\documentclass[useAMS,usenatbib,usegraphicx]{mn2e}
\usepackage[T1]{fontenc}
\usepackage{aecompl}

\usepackage{ txfonts, color, soul,ulem,rotating,subfig,fixltx2e,caption,float,multirow,colortbl} 

    \def\independenT#1#2{\mathrel{\setbox0\hbox{$#1#2$}%
    \copy0\kern-\wd0\mkern4mu\box0}}

\title[000]{Modelling the Hidden Magnetic Field of Low-Mass Stars}
\author[P. Lang et al]
{P. Lang$^{1}$\thanks{E-mail:
pl42@st-andrews.ac.uk},  M. Jardine$^{1}$, J. Morin$^{2,3,4}$, J-F. Donati$^{5}$, S. Jeffers$^{2,6}$, A. A. Vidotto$^{1}$, R. Fares$^{1}$\\
1.SUPA, School of Physics \& Astronomy, University of St Andrews, North Haugh, St Andrews,  KY16 9SS, UK\\
2.Institut f$\ddot{u}$r Astrophysik, Friedrich-Hund-platz 1, 370777 G$\ddot{o}$ttingen\\
3.Dublin Institute for Advanced Studies, School of Cosmic Physics, 31 Fitzwilliam Place, Dublin 2, Ireland\\
4.LUPM, Universit\'{e} Montpellier 2, CNRS, Place Eug\`{e}ne Bataillon, 34095, Montpellier, France\\
5.IRAP-UMR 5277, CNR $\&$ Univ. de Toulouse, 14 Av. E. Belin, F-31400 Toulouse, France\\
6.University of Utrecht, P.O. Box 80000, 3508 TA, Utrecht, The Netherlands\\
}

\begin{document}

\date{Accepted 2014 January 14. Received 2014 January 9; in original form 2013 August 26}

\pagerange{\pageref{firstpage}--\pageref{lastpage}} \pubyear{2014}

\maketitle

\label{firstpage}

\begin{abstract}

Zeeman-Doppler imaging is a spectropolarimetric technique that is used to map the large-scale surface magnetic fields of stars. These maps in turn are used to study the structure of the stars' coronae and winds. This method, however, misses any small-scale magnetic flux whose polarisation signatures cancel out. Measurements of Zeeman broadening show that a large percentage of the surface magnetic flux may be neglected in this way. In this paper we assess the impact of this \textquoteleft missing flux\textquoteright \ on the predicted coronal structure and the possible rates of spin down due to the stellar wind. To do this we create a model for the small-scale field and add this to the Zeeman-Doppler maps of the magnetic fields of a sample of 12 M dwarfs. We extrapolate this combined field and determine the structure of a hydrostatic, isothermal corona. The addition of small-scale surface field produces a carpet of low-lying magnetic loops that covers most of the surface, including the stellar equivalent of solar \textquoteleft coronal holes\textquoteright \  where the large-scale field is opened up by the stellar wind and hence would be X-ray dark. We show that the trend of the X-ray emission measure with rotation rate (the so-called \textquoteleft activity-rotation relation\textquoteright) is unaffected by the addition of small-scale field, when scaled with respect to the large-scale field of each star.  The addition of small-scale field increases the surface flux; however, the large-scale open flux that governs the loss of mass and angular momentum in the wind remains unaffected.  We conclude that spin-down times and mass loss rates calculated from surface magnetograms are unlikely to be significantly influenced by the neglect of small-scale field.

\end{abstract}



\section{Introduction}\label{sec.Introduction}

M dwarfs, much smaller, dimmer and cooler than stars like our Sun, are by far the most common type of star in our galaxy.  The study of these stars has remained limited due to their faintness and in the past it was presumed that M dwarfs were unlikely to host detectable habitable planets.  More recently however, the advantages of searching for habitable planets around M dwarfs have been recognised.  For example, the habitable zone is closer and so it is easier to find planets by radial velocity searches.  Despite the advantages of detecting planets around these stars, M dwarfs have been shown to be extremely magnetically active which may have significant effects on any planetary system.  For example, intense magnetic fields, stellar flares, UV and X-ray emission and the powerful stellar winds \citep{Vidotto_Habitable_2013} may affect planetary atmospheres as well as any potential organisms on these planets.  This makes it vital to investigate how the structure and evolution of the magnetic field, both large-scale and small-scale, can affect coronal properties.

Time-resolved spectropolarimetric observations of a star can be analysed by means of Zeeman Doppler Imaging (ZDI, \citealt{Semel_ZDI_1989}, \citealt{Donati_Sph_harm_2006b}) in order to reconstruct a map of the vector magnetic field on the stellar surface.  ZDI relies on the fact that due to the combination of the properties of the Zeeman effect e.g., rotation-induced Doppler and rotational modulation, a strong relation exists between the distribution of the magnetic field at the surface of a star and the rotational evolution of polarisation in spectral lines during a stellar rotation.  However, several limitations exist: in particular, with the solution being non-unique, a maximum entropy criterion has to be used, and due to the mutual cancellation of polarised signals originating from neighbouring regions of opposite polarities, the maps have a limited spatial resolution and, therefore, systematically miss magnetic flux corresponding to magnetic fields organised on small spatial scales.  The actual resolution is mostly driven by the rotational velocity of the star projected on the observer's line-of-sight (v$\sin i$): the higher the v$\sin i$, the higher the resolution.  In addition, for the inclination of the stellar rotation axis with respect to the line-of-sight differing from 90$^{\circ}$, a part of the star is never visible.  Therefore, in that region there is no constraint on the magnetic field, except that globally it has to satisfy the null-divergence constraint.

Studies based on spectropolarimetric observations and ZDI have provided the first information on the structure of the surface magnetic fields of M dwarfs.  In particular, partly-convective M dwarfs have been shown to host large-scale magnetic fields which are non-axisymmetric and feature a strong toroidal component \citep{Donati_EarlyM_2008}, whereas those close to the limit of full-convection have been shown to host much stronger large-scale field dominated by a mainly axisymmetric poloidal component (e.g., \citealt{Donati_LargeScale_2006a},  \citealt{Morin_MidM_2008},  \citealt{Morin_V374Peg_2008a}).  However, these studies do not constrain the small-scale field component of the magnetic fields of low-mass stars.  In parallel, studies based on the analysis of the Zeeman broadening in unpolarised spectroscopy provide complementary information: the measure of the disc-averaged magnetic field including the contributions of both the large-scale and small-scale components.  \citet{Reiners_Basri_MagneticTopology_2009} compiled measurements of mean magnetic flux from Stokes $I$ (total intensity) and Stokes $V$ (the fractional degree of circular polarisation) parameters for a selection of partially-convective and fully-convective M dwarfs.  They find that the fraction of magnetic flux visible in Stokes $V$ is a small percentage of the total flux measured in Stokes I. This means that a large portion of the magnetic flux stored in magnetic fields is invisible to Stokes $V$.  One possible explanation is that the majority of magnetic flux on M dwarfs is grouped into small structures distributed over the stellar surface, where different polarities cancel each other out in Stokes $V$.  More specifically, \citet{Reiners_Basri_MagneticTopology_2009} find that although for the lower-mass fully-convective stars, the mean magnetic flux does not significantly differ from partially convective stars \citep{ReinersBasri_FirstDirect_2007}, the fraction of the total magnetic flux detected in Stokes V is different for partially-convective and fully-convective stars: 6$\%$ and 14$\%$ respectively.

The aim of this paper is to determine the influence that this small scale field might have on the stellar coronae.  We create a model for small-scale field and add it to the reconstructed surface radial maps for a stellar sample of 12 M Dwarfs \citep{Donati_EarlyM_2008, Morin_MidM_2008} that span the fully-convective boundary.  By comparison with the observed large-scale magnetic field structure, we investigate the effect this small-scale field has on the geometry of the extrapolated 3D magnetic field and subsequent coronal properties, such as open flux, coronal extent, X-ray emission measure and coronal density.  We approach this in two ways: (1) by incorporating small-scale field that has the same surface distribution and magnitude onto each star in the sample; and (2) using the results of \citet{Reiners_Basri_MagneticTopology_2009}, we add in a percentage amount of small-scale field such that the large-scale field contributes only 6$\%$ and 14$\%$ of the total magnetic field, for the partially-convective and fully-convective stars within the sample respectively.

\section{Modelling and Incorporating the small-scale field}\label{sec.Modelling}
\subsection{The Surface Field}

To create small-scale field on the stellar surface we use the synthesised spot brightness maps of \citet{BarnesJeffers_starspot_2011}.  The spots were created using the Doppler imaging code  \textquoteleft Doppler Tomography of Stars\textquoteright \ (DoTS) and all spots were modelled, following \citet{Solanki_starspots_1999}, with circular umbral areas and a ratio of umbral to penumbral area of 1: 3.

We use the spot brightness to allocate field strengths to the centre of the active regions and allow the field strength to fall-off in a Gaussian-like distribution, to the edge of each spot, i.e., 
\begin{equation}
B_{\mathrm{r}}^{\mathrm{ss}} =\frac{B_{\mathrm{max}}} {\Sigma_{\mathrm{brightness}} } e^{- \frac{x^{2}}{2}}        \quad ,
{\label{eq.fieldstrength_1}}
\end{equation}
where $B_{\mathrm{r}}^{\mathrm{ss}}$ represents the field strength in the small-scale field, $\Sigma_{\mathrm{brightness}}$ is the spot brightness, $B_{\mathrm{max}}$ is the arbitrarily chosen maximum field strength, and $x$ is the distance from the centre of the spot.  We note here that a higher spot brightness indicates a lower field strength value.

We impose a condition that the small-scale field must be small enough not to be detected in ZDI i.e., invisible in Stokes $V$.  The typical area over which the circular polarisation cancels out e.g., the area over which the signed magnetic flux cancels or the typical distance between two spots of opposite polarities, corresponds to about 12$^{\circ}$, for a rapidly rotating star with v$\sin i \approx$ 40 $km/s$ e.g., V374 Peg.  This condition means that any active region must have a diameter less than the typical ZDI resolution i.e., $< 5^{\circ}$.  Our synthetic maps assume spots with radii $\le$1$^{\circ}$.  

To ensure the small-scale field is evenly distributed over the entire surface of the star, we keep the spot coverage constant.  We, therefore, find that an appropriate parameter to vary in the model is $B_{\mathrm{max}}$.  Taking into account the magnitude of the field detected in ZDI for our sample of partly-convective and fully convective M Dwarfs, we (1) fix the value of $B_{\mathrm{max}}$ to be either $\pm$500$G$ or $\pm$1000$G$; and (2) set the value of $B_{\mathrm{max}}$ such that the large-scale field contributes between approximately 6$\%$ and 14$\%$ of the total field respectively, as indicated in \citet{Reiners_Basri_MagneticTopology_2009}.  The values for the average radial flux in each case can be found in Table (\ref{tab.stellardata_dipole_contribution}).

\begin{figure}
	\begin{center}
	{\label{fig.48percent} \includegraphics[width=80mm]{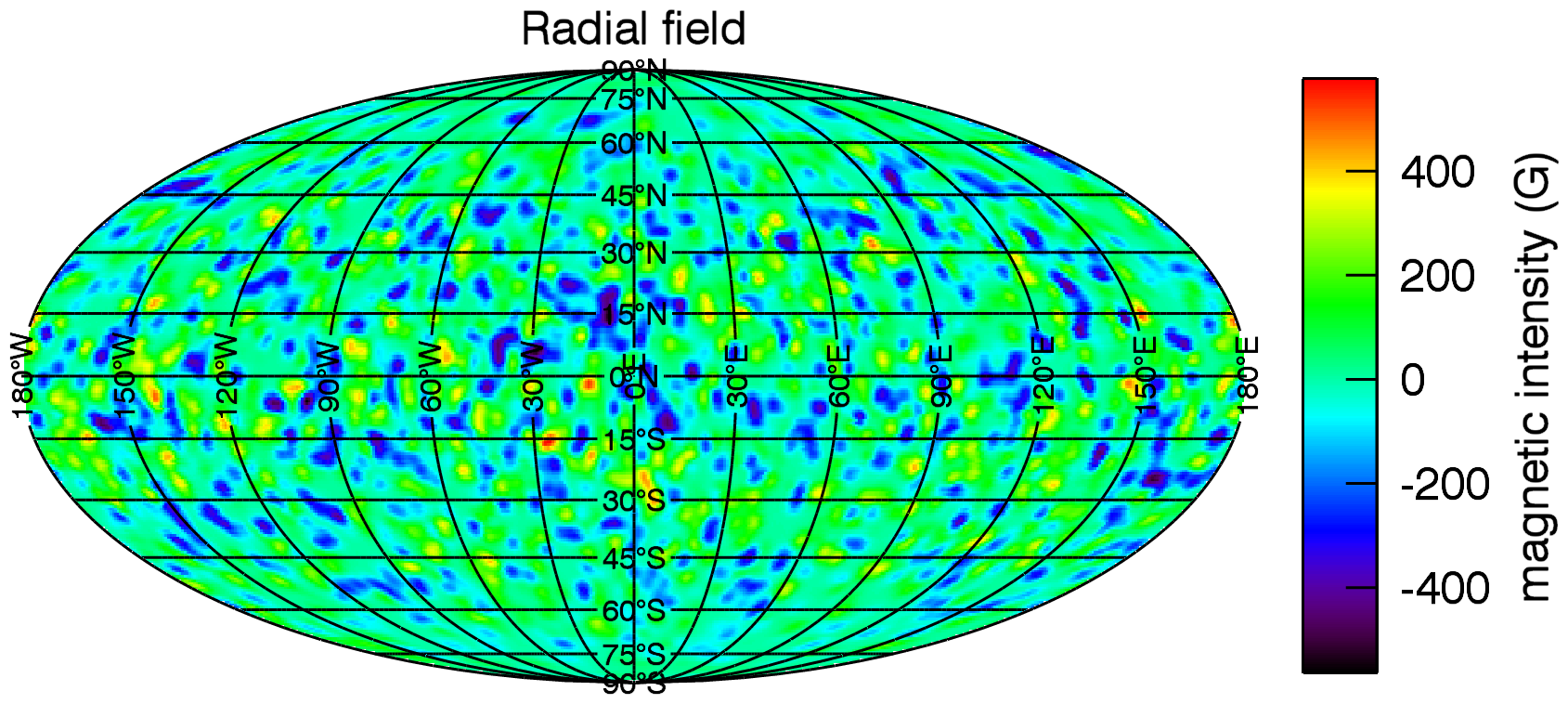}}	
	\caption{Radial magnetic surface map for the model of small scale field covering 62$\%$ of the stellar surface.  $|B_{\mathrm{max}}|$ is 500G with an unsigned surface flux value $\phi_{\mathrm{surface}} \approx 10^{24}$Mx.  
	}	\label{fig.Spot_distribution}
	\end{center}
\end{figure}

The surface magnetic radial map for the small-scale field is shown in Fig. (\ref{fig.Spot_distribution}) where the spot distribution covers approximately 62$\%$ of a model star.  Fig. (\ref{fig.Spot_distribution}) represents the case where $B_{\mathrm{max}}$ = 500G.  The simulated magnetic radial maps for the small-scale field are added to the reconstructed radial maps obtained through ZDI and new surface maps with both large-scale and small-scale field are created for each M dwarf i.e., $B_{\mathrm{Total}}$ = $B_{\mathrm{r}}^{\mathrm{ss}} + B_{\mathrm{r}}^{\mathrm{ls}} $.  

\subsection{The Coronal Field}\label{sec.FieldExtrapolation}

The magnetic field is extrapolated above the stellar surface using the potential-field source surface (PFSS) method \citep{Altschuler_Newkirk_1969}, where the magnetic field is assumed to be current-free ($\underline \nabla \times \underline B = 0$) and divergence free ($\underline \nabla \cdot \underline B = 0$).  In a format similar to \citet{Jardine_ABDor_1999} the components for the coronal magnetic field are determined from the solution to Laplace's equation $\nabla ^2 \psi = 0$, where $\psi$ is the scalar potential:
\begin{equation}
B_{r}=-\displaystyle\sum_{l=1}^{N}\displaystyle\sum_{m=1}^{l}[la_{lm}r^{l-1} - (l+1)b_{lm}r^{-(l+2)}]P_{lm}(\cos\theta)e^{im\phi}     
\label{eq.field_1}
\end{equation}
\begin{equation}
B_{\theta}=-\displaystyle\sum_{l=1}^{N}\displaystyle\sum_{m=1}^{l}[a_{lm}r^{l-1} + b_{lm}r^{-(l+2)}]\frac{d}{d\theta}P_{lm}(\cos\theta)e^{im\phi}
\label{eq.field_2}
\end{equation}
\begin{equation}
B_{\phi}=-\displaystyle\sum_{l=1}^{N}\displaystyle\sum_{m=1}^{l}[a_{lm}r^{l-1} + b_{lm}r^{-(l+2)}]P_{lm}(\cos\theta)\frac{im}{\sin\theta}e^{im\phi}
\label{eq.field_3}
\end{equation}
with $B_{r}, B_{\theta}, B_{\phi}$ representing the radial, meridional and azimuthal components of the magnetic field, respectively, $\mathrm{P_{lm}}$ represents the associated Legendre polynomials, $a_{lm}$ and $b_{lm}$ are the amplitudes of the spherical harmonics, \textit{l} is the spherical harmonic degree, \textit{m} is the order or azimuthal number and $r = R/R_{\star}$.

To extrapolate the 3D coronal field and determine the amplitude of the spherical harmonics, $a_{lm}$ and $b_{lm}$, we apply two boundary conditions.  The upper condition is that at the Source Surface, $R_{ss} \ $\citep{Schatten_SourceSurface_1969}, the field opens and is purely radial ($B_\theta=B_\phi=0$), while the lower boundary condition imposes the observed radial field.  We choose the solar value for the source surface at 2.5$R_{*}$.  The code used to extrapolate the field is a modified version of the global diffusion model developed by \citet{VanBallegooijen_Diffusion_1998}.

\begin{figure}
	\begin{center}
	\includegraphics[width=45mm]{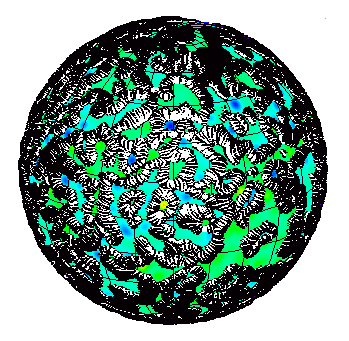}
	\includegraphics[width=19mm]{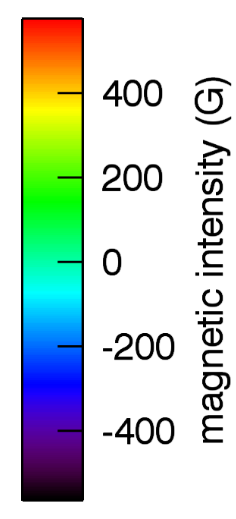}\\
\caption{3D coronal extrapolation of the small-scale field shown in Fig. \ref{fig.Spot_distribution}.  Colours are scaled to the maximum and minimum values of the surface radial magnetic field component.
	 \label{fig.SmallScaleField} }
	 	\end{center}
\end{figure}
The extrapolated 3D small-scale field is shown in Fig. (\ref{fig.SmallScaleField}), where the field lines remain closed and close to the stellar surface.  This extrapolation demonstrates that the small-scale field produces a \textquoteleft carpet\textquoteright \ of low-lying loops across the surface.  

\subsection{X-ray Emission Model}\label{sec.EmissionMeasure}

The structure of the magnetic field is influenced by the strength of the surface field and the distribution of plasma.  Following the model in \citet{Lang_large-scale_2012}, the density structure can be estimated for the extrapolated corona by assuming the plasma is hydrostatic and isothermal and that the gas pressure at the stellar surface is proportional to the magnetic pressure ($p_{o} = \kappa B_{o}^{2}$).  $\kappa$ is a constant of proportionality relating the base gas pressure $p_{o}$ to magnetic pressure $B_{o}$ through the magnetic constant $2\mu$.  The value of $\kappa$ is chosen such that the coronal densities lie within the observed range for M Dwarfs, $10^{9}-10^{12}\mathrm{cm^{-3}}$ \citep{Ness_Density_ADLeo_2002,Ness_Density_2004}.  Typical values for $\log\kappa = [ -5 : -7]$

We assume that the pressure varies along each field line according to
\begin{equation}
p=p_{o}e^{\int{\frac{\underline g \cdot \underline B    ds}{|B|}}}           \quad , 
\label{eq.pressure_1}
\end{equation}
as described by \citet{Jardine_XrayCorona_2002} and \citet{Gregory_XrayRotMod_2006}.  Expanding the (dimensionless) component of gravity along the field line ($\underline g \cdot \underline B$), we have 
\begin{equation}
p=\kappa B_{o}^{2} exp\left[\int {\frac{\left(\frac{-\phi_{g}}{r^{2}} + \phi_{c} r \sin^{2}\theta \right)B_r + \left(\phi_{c} r \sin\theta \cos\theta \right)B_{\theta}} {\sqrt{B_{r}^{2}+B_{\theta}^{2}+B_{\phi}^{2}}}} \mathrm{d}s \right]     \quad,
\label{eq.pressure}
\end{equation}
Where $r = R/R_{\star}$ and the ratios of centrifugal ($\phi_{c}$) and gravitational ($\phi_{g}$) to thermal energy are given by
\begin{equation}
\phi_{c}=m_{e} \left( \frac{(\omega R_{\star})^{2}}{k_{B} T} \right)
\label{eq.cent-thermal}
\end{equation}
\begin{equation}
\phi_{g}=m_{e} \left( \frac{G M_{\star}}{R_{\star}k_{B} T} \right) \quad ,
\label{eq.grav-thermal}
\end{equation} 
where $R_{\star}$ is the stellar radius, $M_{\star}$ is the stellar mass, $\omega$ is the stellar rotation rate, $k_{B}$ is the Boltzmann constant, $G$ is the gravitational constant and $m_{e}$ is the electron mass.

To ensure that only regions of the closed stellar corona contribute towards the emission measure, the gas pressure along open field lines is taken to be zero.  In addition to this, if there is any over-pressure along the designated closed loops i.e. gas pressure ($p=2n_{e}kT$) $\ge$ magnetic pressure ($p_{B}=B^{2}/2\mu$), then the pressure at that grid point is also set to zero. 

Assuming the gas is optically thin, the X-ray emission measure varies with density, i.e.
\begin{equation}
EM(r) = \int n_e^{2}   \mathrm{d}V      \quad .
\label{eq.EmissionMeasure2}
\end{equation}  

The temperature ($\mathrm{T = 2\times10^{6}}$K), source surface ($R_{ss}$=$2.5R_{*}$) and constant of proportionality $\kappa$ ($10^{-6}$) are kept constant in this paper (for more details see \citet{Lang_large-scale_2012}).  

\section{Results}\label{sec.results}

\subsection{Field Structure}\label{sec.results_structure}

\begin{figure*}
	\begin{center}
	\subfloat[Large-scale radial field at stellar surface]{\label{fig.GJ 49-surface}\includegraphics[width=70mm]{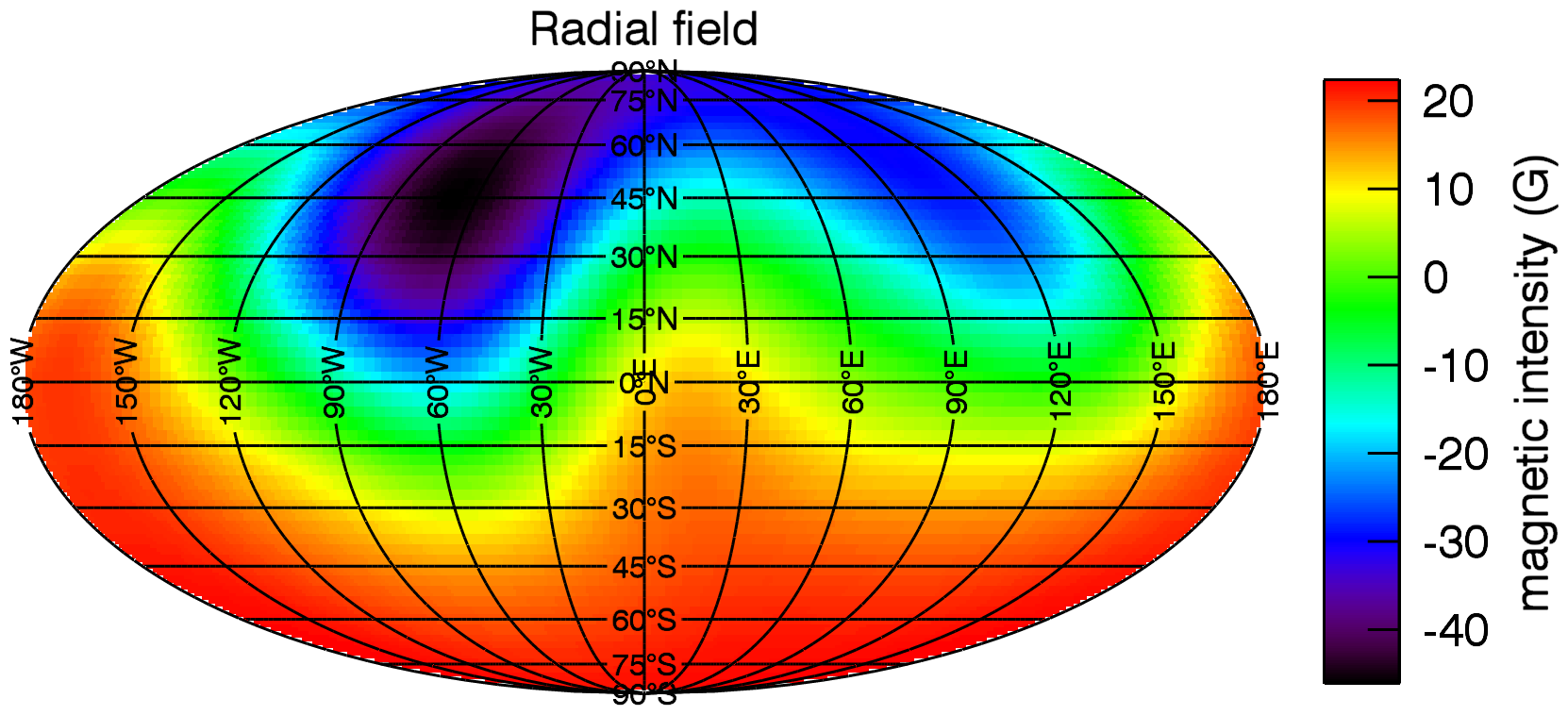}}
	\subfloat[Large-scale radial field at source surface]{\label{fig.GJ 49-surface-b}\includegraphics[width=70mm]{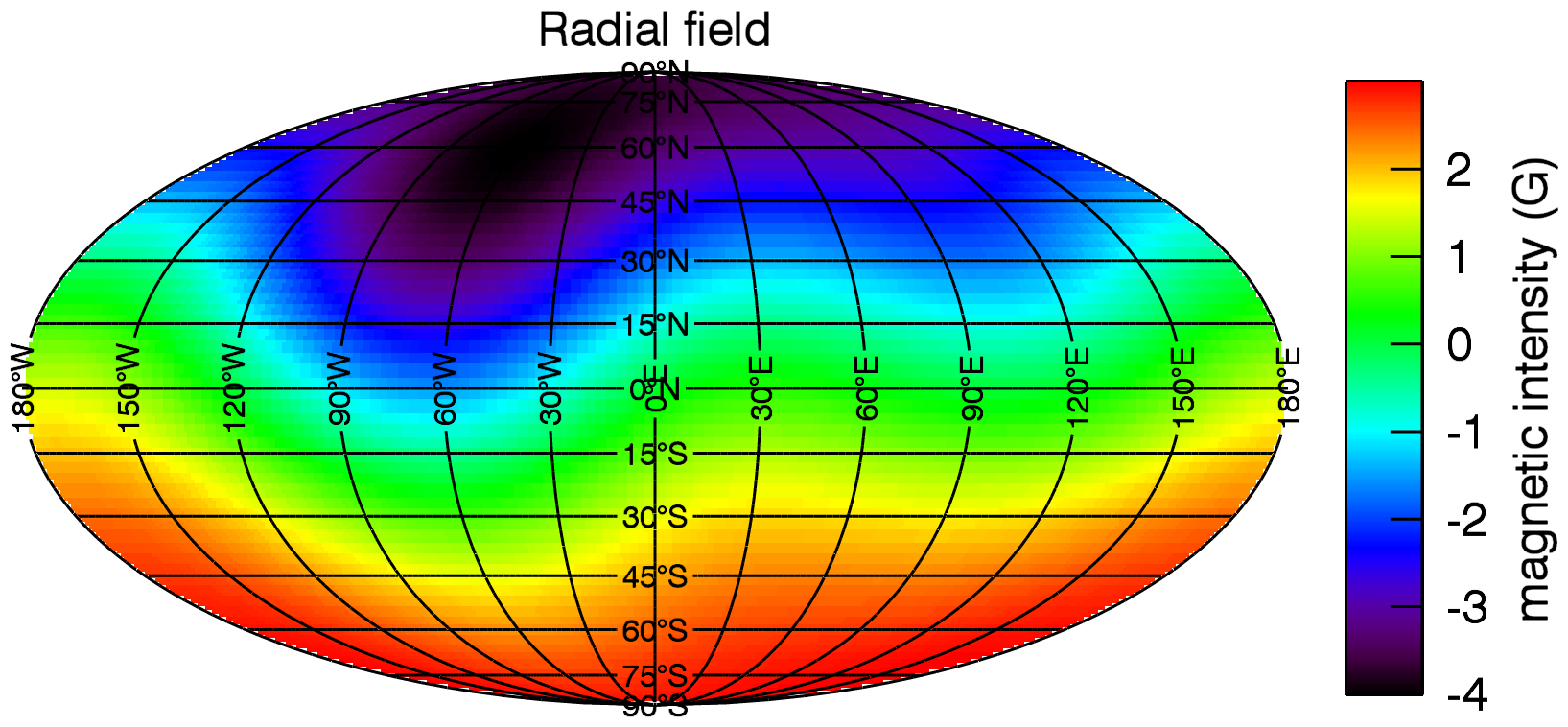}}\\
	\subfloat[Large- + small-scale radial field at stellar surface]{\label{fig.GJ 49-surface-c}\includegraphics[width=70mm]{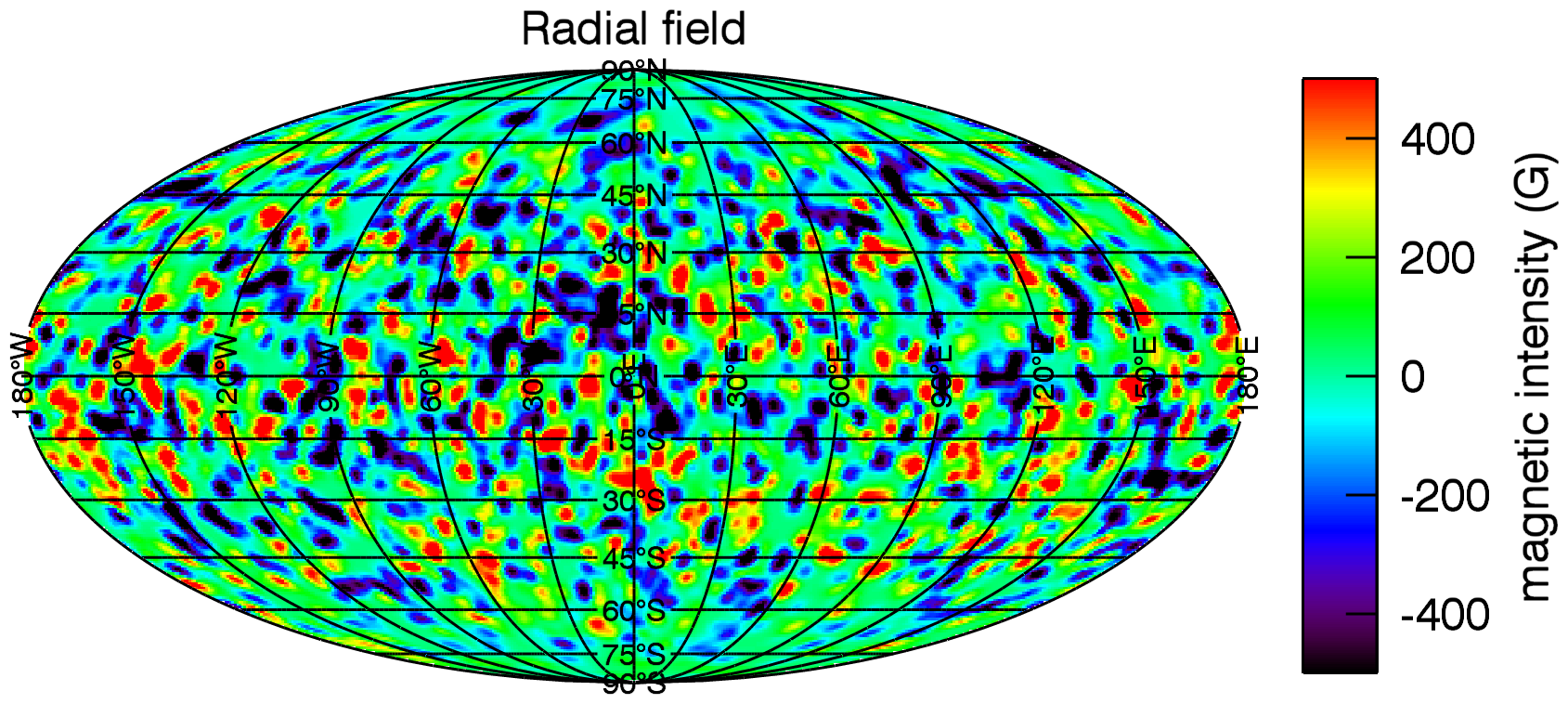}}
	\subfloat[Large- + small-scale radial field at source surface]{\label{fig.GJ 49-surface-d}\includegraphics[width=70mm]{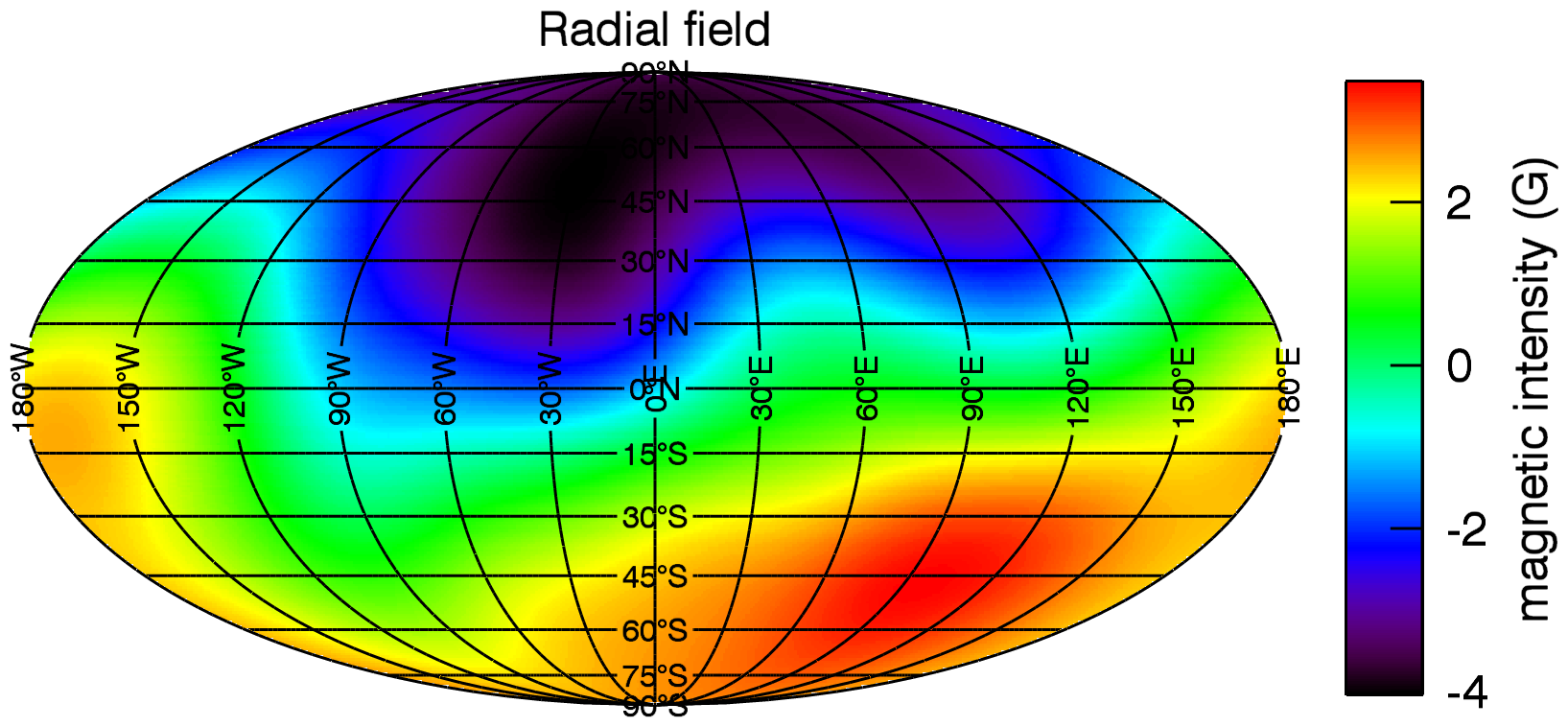}}
	\caption{\textit{Upper:} Large-scale (ZDI) reconstructed radial field maps for GJ 49 at (a) the stellar surface, and (b) the source surface.  \textit{Lower:} ZDI + small-scale radial field map for GJ 49 at (c) the stellar surface, and (d) the source surface.  Small-scale field is scaled according to the results of \citet{Reiners_Basri_MagneticTopology_2009} such that $\mathrm{B_{ls} = 6\%B_{Total}}$, for partly-convective M dwarfs.}
	\label{fig.surface_field}
	\end{center}
\end{figure*}

With the addition of small-scale field $\mathrm{B}$ drops more rapidly with height, close to the stellar surface.  As such, we do not find any great change in the large-scale field structure.  This is evident from Fig. (\ref{fig.surface_field}) which shows the radial field at both the stellar surface and the source surface.  Fig. (\ref{fig.GJ 49-surface}) $\&$ (\ref{fig.GJ 49-surface-c}) which represent the large-scale and large- + small-scale field at the stellar surface respectively, show very different topologies; however, when this is extrapolated out to the source surface (Fig. (\ref{fig.GJ 49-surface-b}) $\&$ (\ref{fig.GJ 49-surface-d})) the topologies are similar.  We conclude from this that the magnetic pressure falls off with radius more quickly with the addition of small-scale field leaving only the large-scale components near the source surface.

Fig. (\ref{fig.surface_get_1}) shows the coronal magnetic field of; (1) the large-scale field extrapolated from the reconstructed radial maps (left-hand column); and (2) the large- + small-scale field, where the small-scale field is scaled according to \citet{Reiners_Basri_MagneticTopology_2009}, such that $B_{\mathrm{ls}} = 6\%B_{\mathrm{Total}}$, for partly-convective M dwarfs and $B_{\mathrm{ls}} = 14\%B_{\mathrm{Total}}$, for fully-convective M dwarfs (right-hand column).  A comparison of the extrapolations in both the left- and right-hand columns from Fig. (\ref{fig.surface_get_1}) show the location of coronal holes where the stellar wind is emitted and the angle of the magnetic dipole axis from the rotation pole remain largely unchanged on the majority of the stars in the sample.  However, the extrapolations of DT Vir, OT Ser and EQ Peg A show small changes in the structure of the closed large-scale field.  

\begin{figure*}
	\begin{center}
	\subfloat[GJ 182 (07): 0.75$M_{\odot}$; \ Large-scale field]{\label{fig.GJ 182-a}\includegraphics[width=45mm]{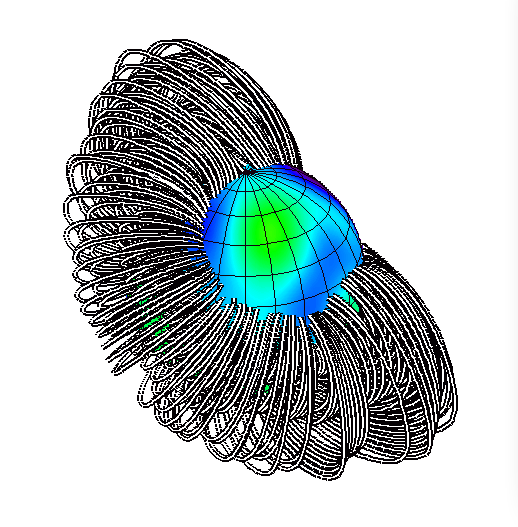}}
	\includegraphics[width=15mm]{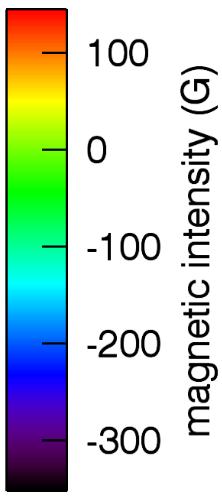}
	\subfloat[Large- + small-scale field]{\label{fig.GJ 182-c}	\includegraphics[width=45mm]{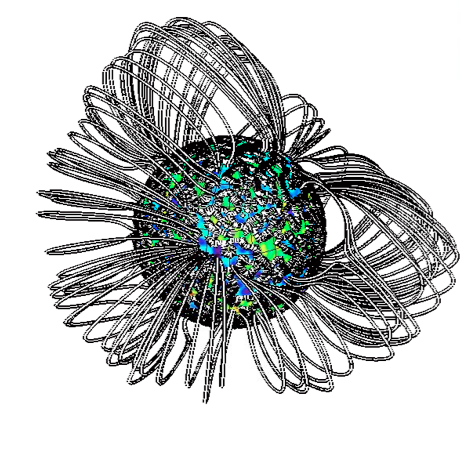}}
	\includegraphics[width=17mm]{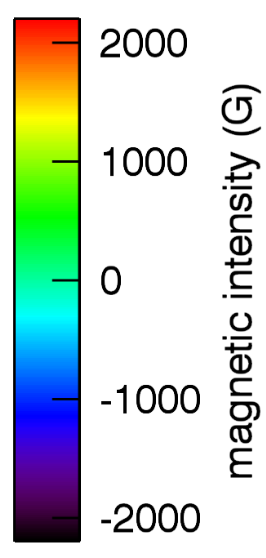}

	\subfloat[DT Vir (08): 0.59$M_{\odot}$;  \ Large-scale field]{\label{fig.DTVir-a} \includegraphics[width=45mm]{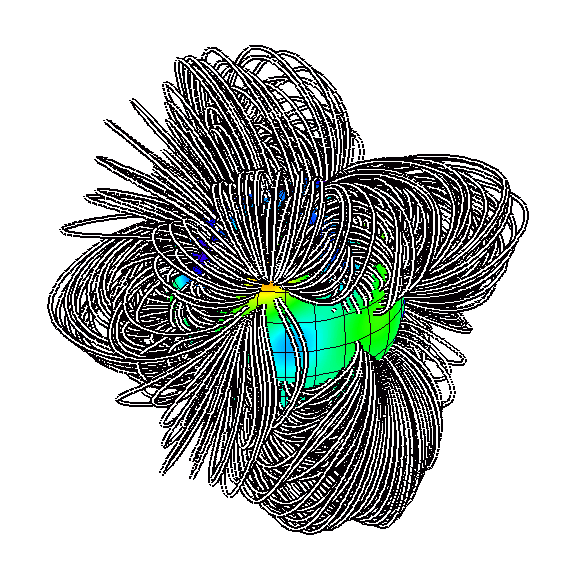}}
	\includegraphics[width=15mm]{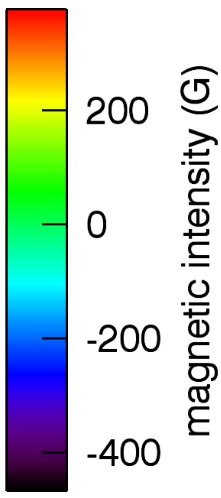}
	\subfloat[Large- + small-scale field]{\label{fig.DTVir-c}	\includegraphics[width=45mm]{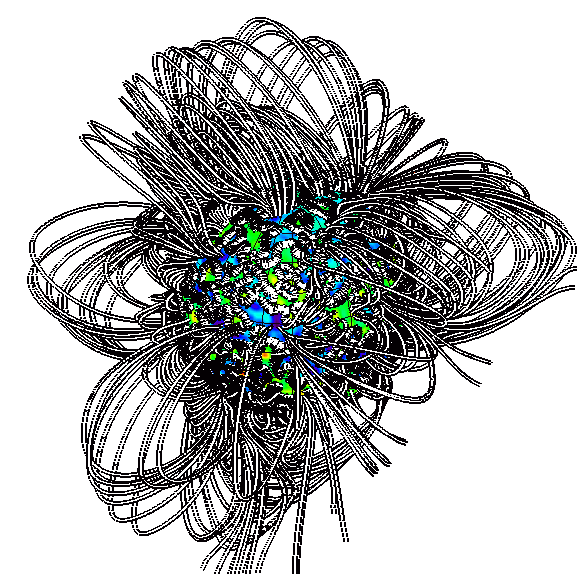}}
	\includegraphics[width=17mm]{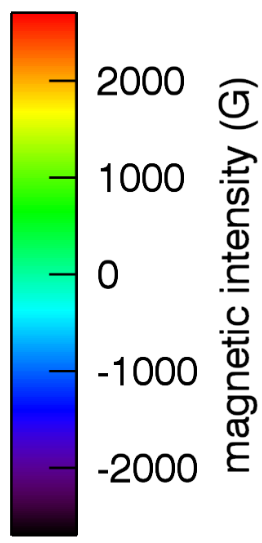}

	\subfloat[DS Leo (08): 0.58$M_{\odot}$; \ Large-scale field]{\label{fig.DSLeo-a} \includegraphics[width=45mm]{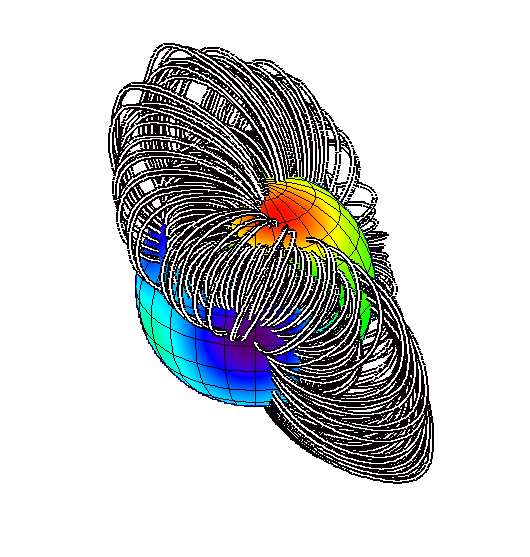}}
	\includegraphics[width=15mm]{ColourBars/LargeScale/m_1008_Bv_colour.png}
	\subfloat[Large- + small-scale field]{\label{fig.DSLeo-c}	\includegraphics[width=45mm]{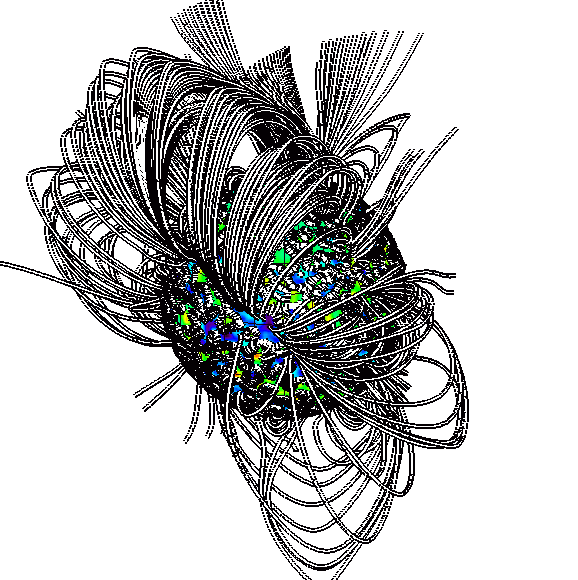}}
	\includegraphics[width=17mm]{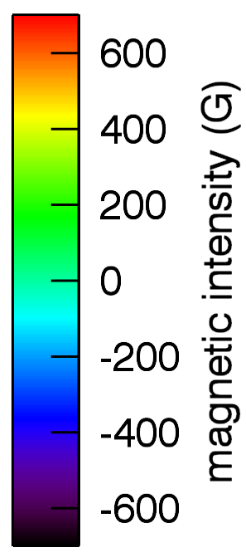}

	\subfloat[GJ 49 (07) : 0.57$M_{\odot}$; \ Large-scale field]{\label{fig.GJ49-a} \includegraphics[width=45mm]{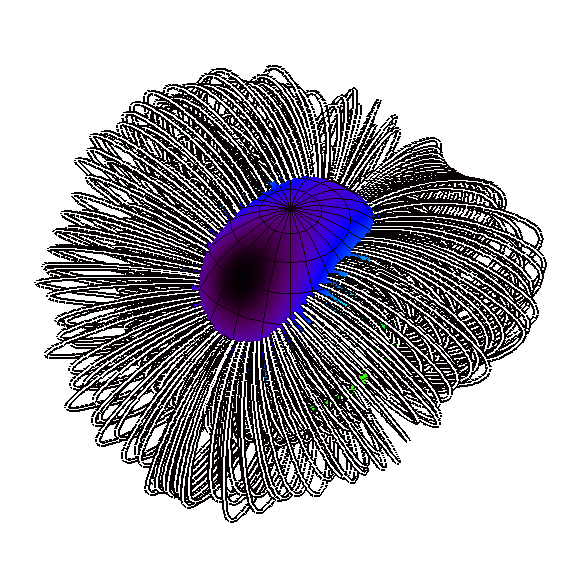}}
	\includegraphics[width=15mm]{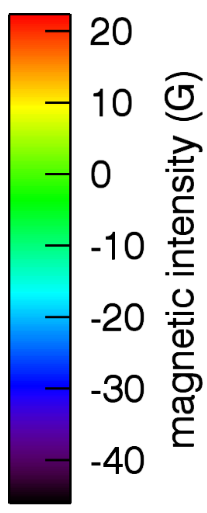}
	\subfloat[Large- + small-scale field]{\label{fig.GJ49-c}	\includegraphics[width=45mm]{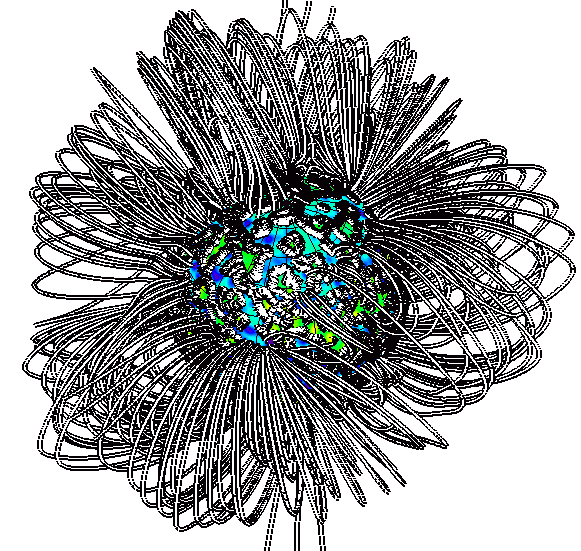}}
	\includegraphics[width=17mm]{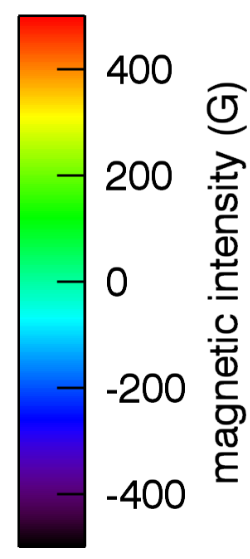}

	\caption{Column (1) shows the 3D coronal extrapolation for the reconstructed surface radial maps for our sample of M dwarfs.  Column (2) is the extrapolation for case (2) with the combination of the small-scale field scaled with respect to the large-scale field e.g., $B_{\mathrm{ls}} = 6\%B_{\mathrm{Total}}$, for the partly-convective M dwarfs and $B_{\mathrm{ls}} = 14\%B_{\mathrm{Total}}$, for the fully convective M dwarfs.}
	\label{fig.surface_get_1}
	\end{center}
\end{figure*}

\begin{figure*}
	\begin{center}

	\subfloat[OT Ser (08): 0.55$M_{\odot}$; \ Large-scale field]{\label{fig.OTSer-a} \includegraphics[width=45mm]{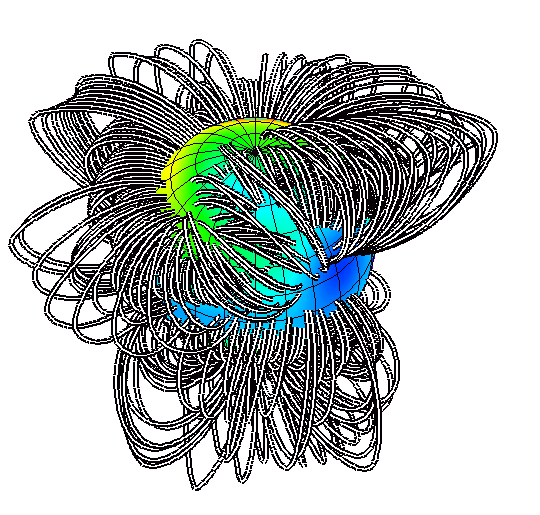}}
	\includegraphics[width=15mm]{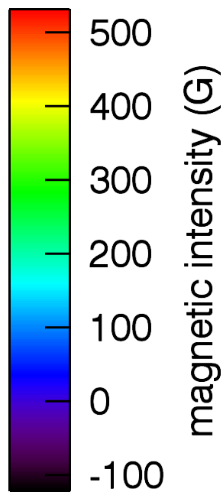}
	\subfloat[Large- + small-scale field]{\label{fig.OTSer-c}	\includegraphics[width=45mm]{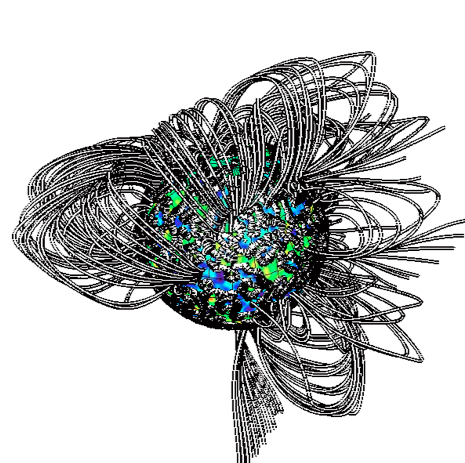}}
	\includegraphics[width=17mm]{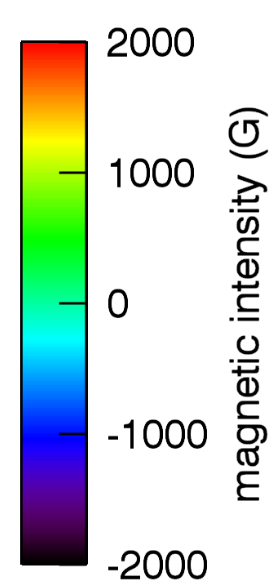}

	\subfloat[CE Boo (08): 0.48$M_{\odot}$; \ Large-scale field]{\label{fig.CEBoo-a} \includegraphics[width=45mm]{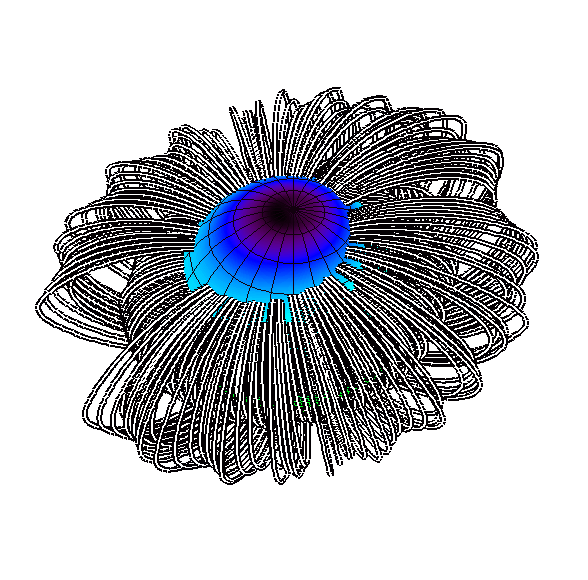}}
	\includegraphics[width=15mm]{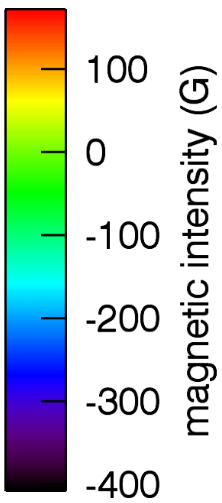}
	\subfloat[Large- + small-scale field]{\label{fig.CEBoo-c}	\includegraphics[width=45mm]{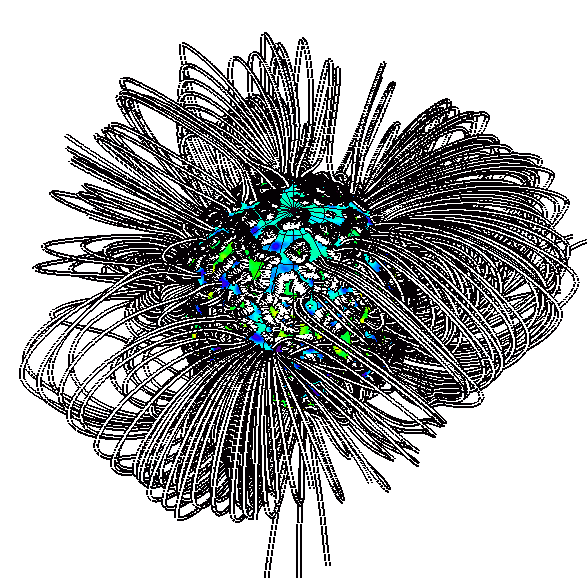}}
	\includegraphics[width=17mm]{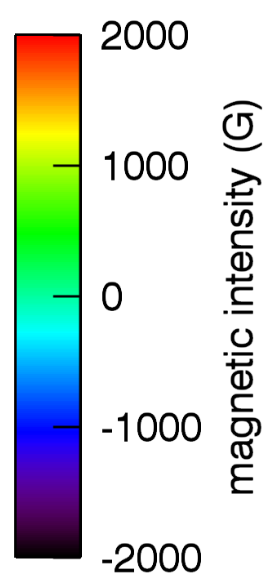}

	\subfloat[AD Leo (07): 0.42$M_{\odot}$; \ Large-scale field]{\label{fig.ADLeo-a} \includegraphics[width=45mm]{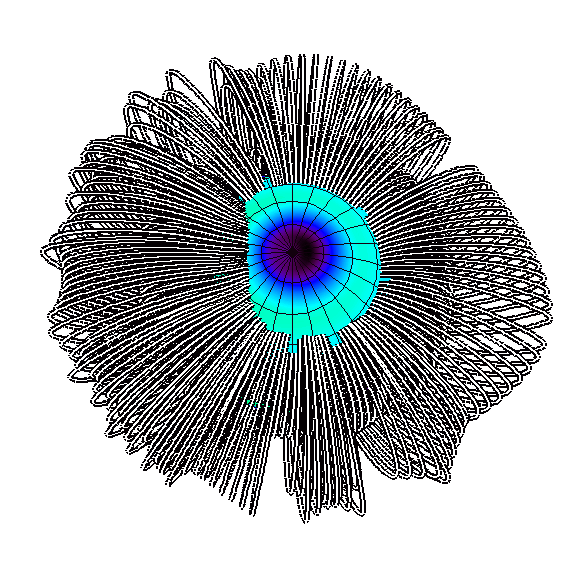}}
	\includegraphics[width=15mm]{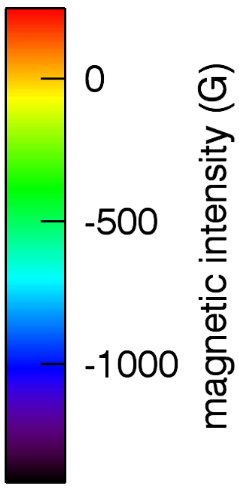}
	\subfloat[Large- + small-scale field]{\label{fig.ADLeo-c}	\includegraphics[width=45mm]{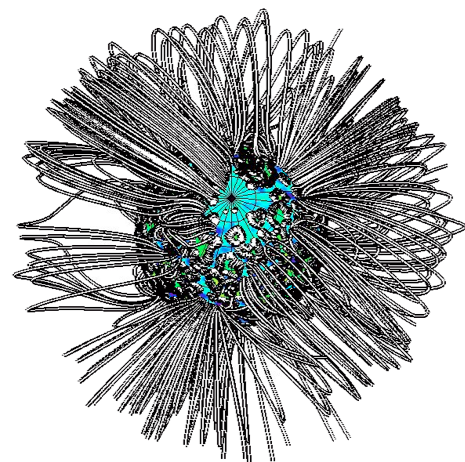}}
	\includegraphics[width=17mm]{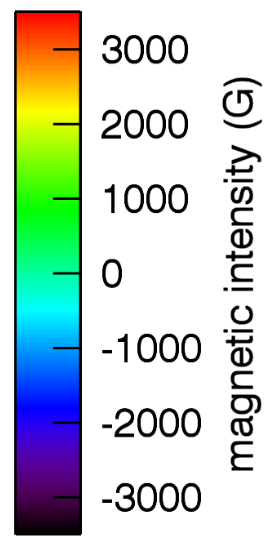}

	\subfloat[EQ Peg A (06): 0.39$M_{\odot}$; \ Large-scale field]{\label{fig.EQPegA-a} \includegraphics[width=45mm]{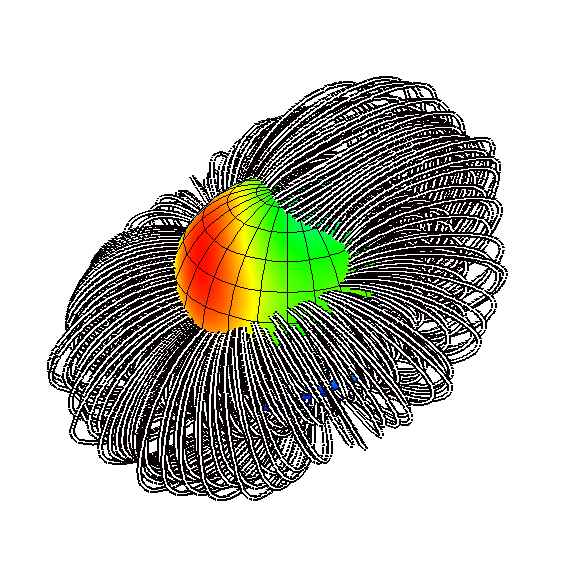}}
	\includegraphics[width=15mm]{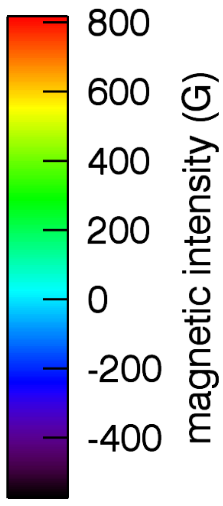}
	\subfloat[Large- + small-scale field]{\label{fig.EQPegA-c}	\includegraphics[width=45mm]{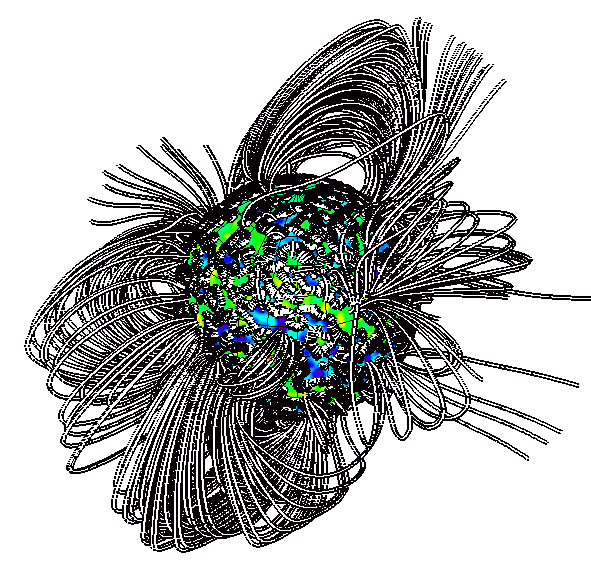}}
	\includegraphics[width=17mm]{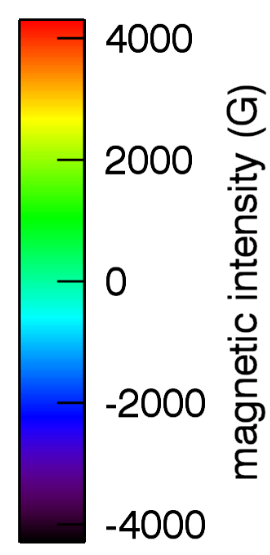}

	\contcaption{}	\label{fig.surface_get_2}
	\end{center}
\end{figure*}

\begin{figure*}
	\begin{center}
	
	\subfloat[EV Lac (06): 0.32$M_{\odot}$; \ Large-scale field]{\label{fig.EVLac-a} \includegraphics[width=45mm]{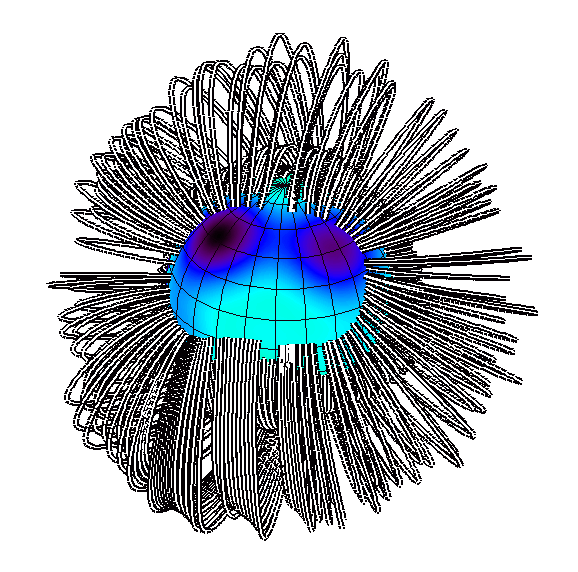}}
	\includegraphics[width=15mm]{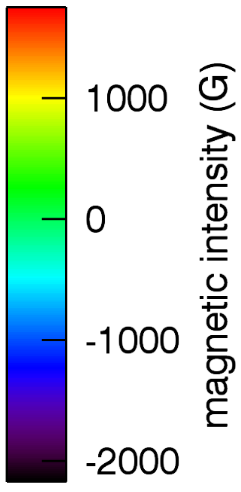}
	\subfloat[Large- + small-scale field]{\label{fig.EVLac-c}	\includegraphics[width=45mm]{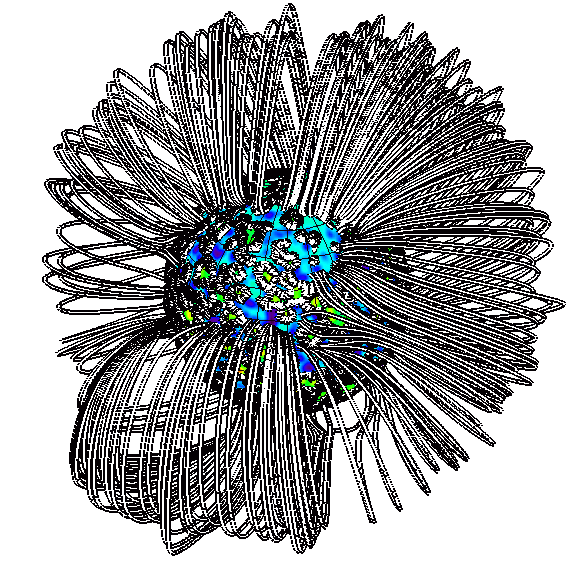}}
	\includegraphics[width=17mm]{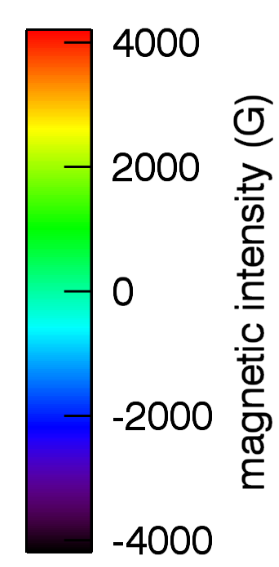}

	\subfloat[YZ CMi (07): 0.31$M_{\odot}$; \ Large-scale field]{\label{fig.YZCMi-a}\includegraphics[width=45mm]{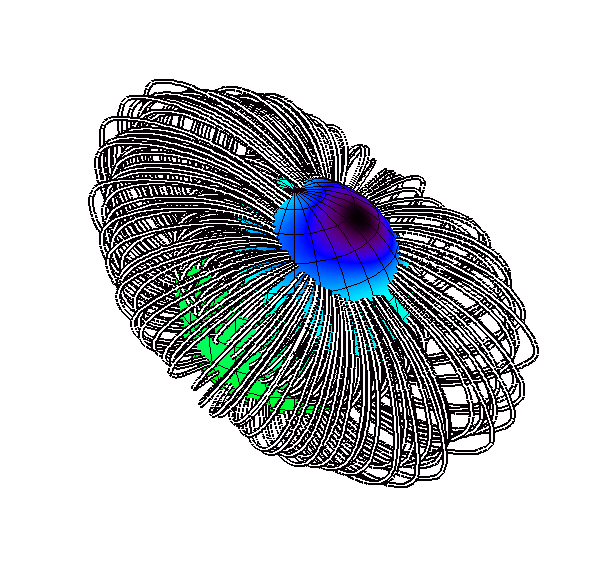}}
	\includegraphics[width=15mm]{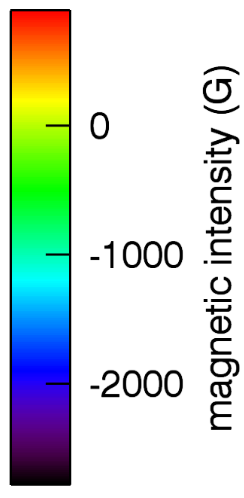}
	\subfloat[Large- + small-scale field]{\label{fig.YZCMi-c}	\includegraphics[width=45mm]{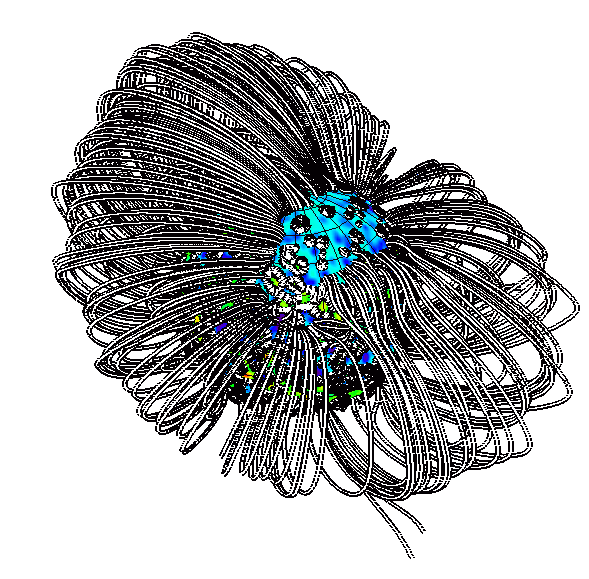}}
	\includegraphics[width=17mm]{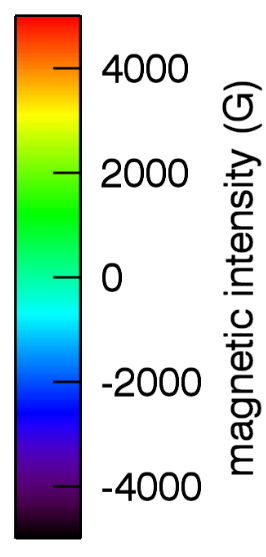}

	\subfloat[V374 Peg (05/06): 0.28$M_{\odot}$; \ Large-scale field]{\label{fig.V374-a} \includegraphics[width=45mm]{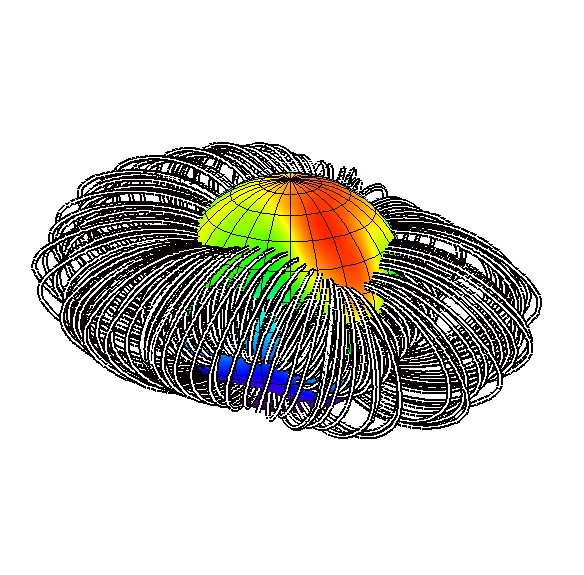}}
	\includegraphics[width=15mm]{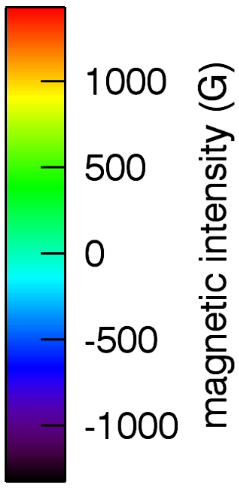}
	\subfloat[Large- + small-scale field]{\label{fig.V374-c} \includegraphics[width=45mm]{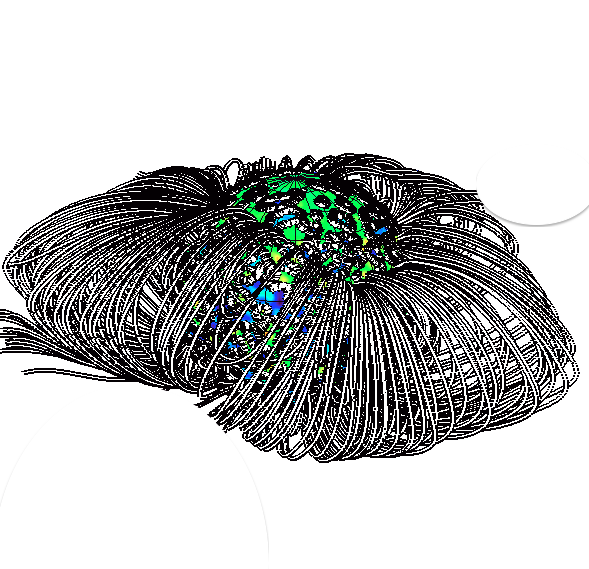}}
	\includegraphics[width=17mm]{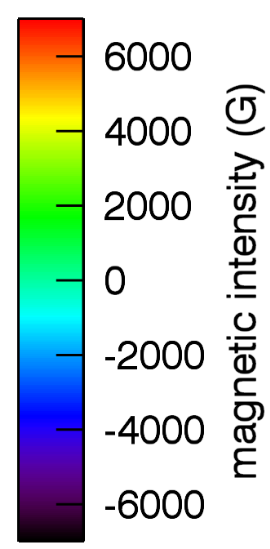}

	\subfloat[EQ Peg B (06): 0.25$M_{\odot}$; \ Large-scale field]{\label{fig.EQPegB-a} \includegraphics[width=45mm]{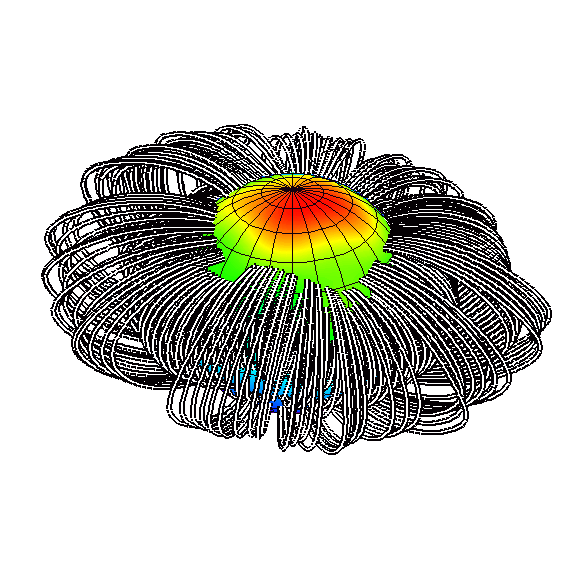}}
	\includegraphics[width=15mm]{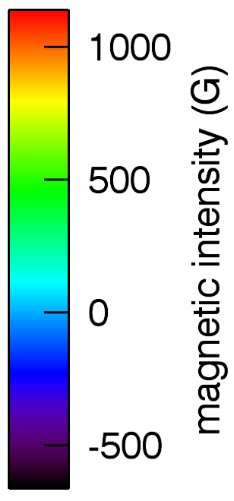}
	\subfloat[Large- + small-scale field]{\label{fig.EQPegB-c}	\includegraphics[width=45mm]{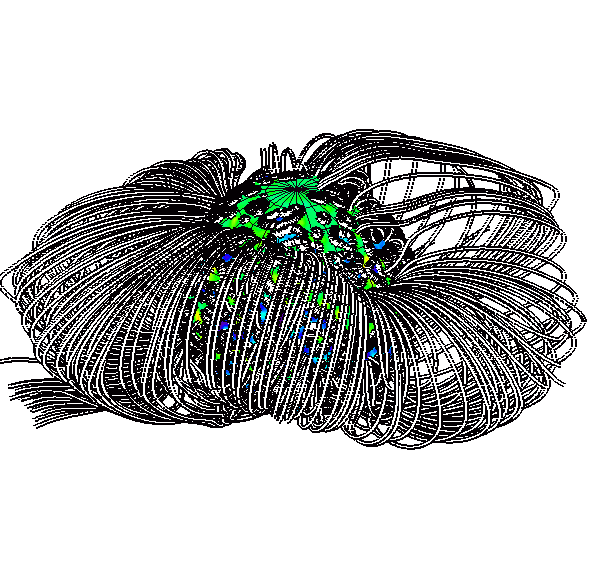}}
	\includegraphics[width=17mm]{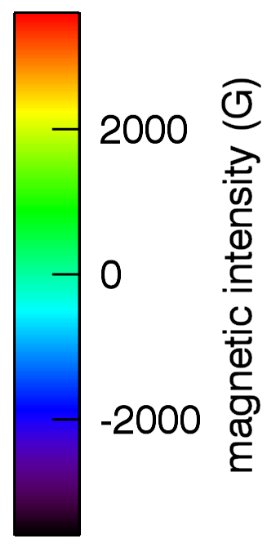}

	\contcaption{}	\label{fig.surface_get_4}
	\end{center}
\end{figure*}

\subsection{Activity-Rotation Relation}

\begin{table*}
\caption{Results for the coronal properties for the sample of M dwarfs.  The predicted values for the logarithmic emission measure (both magnitude, $\log{EM}$, and rotational modulation, Rot Mod) and logarithmic coronal density, $\log{\overline n_{e}}$, for the observed large-scale field, are from \citet{Lang_large-scale_2012}.   The logarithmic emission measure (both magnitude, $\log{\mathrm{EM}}$, and rotational modulation, Rot Mod) and logarithmic coronal density, $\log{\overline n_{e}}$ for the simulated large-scale + small scale field at $B_{\mathrm{max}}$ = 500G and $B_{\mathrm{max}}$ = 1000G as well as for $B_{\mathrm{ls}} = 6\% B_{\mathrm{Total}}$ for partly-convective stars and $B_{\mathrm{ls}} = 14\% B_{\mathrm{Total}}$ for fully convective stars, are from this work.   \label{tab.stellardata}}
\centering
\begin{tabular}{!{\color{black}\vrule}lc!{\color{black}\vrule}ccc!{\color{black}\vrule}ccc!{\color{black}\vrule}ccc!{\color{black}\vrule}ccc!{\color{black}\vrule}ccc!{\color{black}\vrule}}
\hline
\multirow{3}{*} & & & Large-scale & & & + 500G & & & + 1000G & & & + $\%$& \\
\hline
Star & Sp Type &$\log{EM}$ &Rot Mod &$\log{\overline n_{e}}$ &$\log{EM}$ &Rot Mod &$\log{\overline n_{e}}$&$\log{EM}$ &Rot Mod &$\log{\overline n_{e}}$ &$\log{EM}$ &Rot Mod &$\log{\overline n_{e}}$\\
& & ($cm^{-3}$) & $\%$ & ($cm^{-3}$) & $cm^{-3}$ & $\%$ &($cm^{-3}$)& $cm^{-3}$ & $\%$ &($cm^{-3}$) & $cm^{-3}$ & $\%$ &($cm^{-3}$)\\
\hline
\hline
GJ 182 & M0.5 & 50.3&12&8.6&51.2&31&9.2&52.1&31&9.8& 55.4 & 25 & 11.0 \\
DT Vir & M0.5  &50.9&3& 9.3&50.4&21&9.1&51.6&10&9.8& 54.8 & 20 & 11.3 \\
DS Leo & M0  & 48.4&10&7.9&49.9&30&9.1& 50.8 & 43 & 9.9 & 53.6 & 18 & 10.7 \\
GJ 49 & M1.5  & 46.8&18&7.1&50.0&24&9.2& 50.9 & 26 & 9.8 & 52.0 & 19 & 9.9 \\
OT Ser & M1.5 &50.7& 49&9.2&50.8&24&9.3&51.6&13&9.8& 54.2 & 20 & 11.3 \\
CE Boo & M2.5 & 49.5&2&8.5&50.2&4&9.2&51.3&10&9.7& 54.9 & 20 & 11.4\\
AD Leo & M3 & 50.2&3&8.9&50.6&0.4&9.4&50.6&0.4&9.4& 54.5 & 10 & 11.7 \\
EQ Peg A & M3.5 & 51.4&51&9.7&51.5&44&9.7&51.8&48&10.0& 54.9 & 27 & 12.0 \\
EV Lac & M3.5 &52.0&12&10.1&52.1&4&10.2&52.2&10&10.3& 55.4 & 6 & 12.2\\
YZ CMi & M4  &50.2&10& 10.3&52.4&10&10.4&52.4&9&10.4& 55.7 & 28 & 12.1 \\      
V374 Peg & M4 &52.6&27&10.3&52.9&22&10.4&52.8&24&10.4& 56.6 & 22 &12.4 \\
EQ Peg B & M4.5 & 51.1&14&9.6&51.3&12&9.8&51.6&21&10.1& 54.3 & 22 & 11.8\\
\hline
\end{tabular}
\end{table*}

\begin{figure}
	\includegraphics[width=80mm]{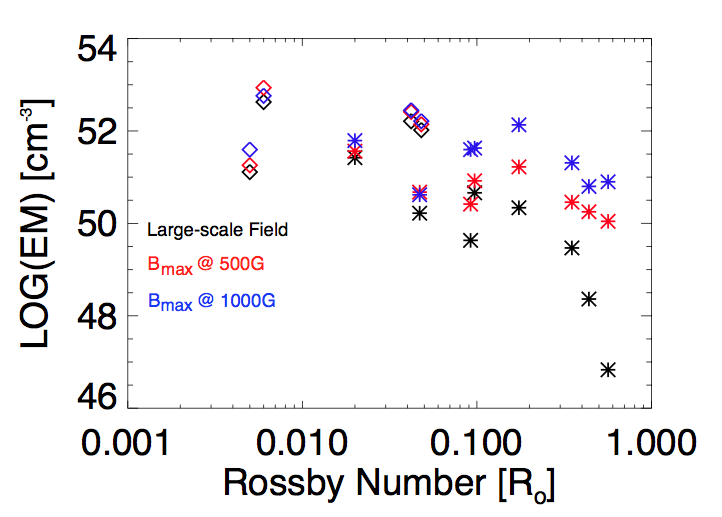}
\caption{X-ray Emission Measure as a function of Rossby number for both the large-scale field (black symbols) and the simulated small-scale + large-scale field ($B_{\mathrm{max}}$ = 500G red symbols, $B_{\mathrm{max}}$ = 1000G blue symbols).  \textit{Symbols}: asterisks represent partly convective dwarfs, $M > 0.4M_{\odot}$, and diamond represent fully convective dwarfs, $M \le 0.4M_{\odot}$.
	 \label{fig.EM_RossbyRatio6_Early_Late} }
\end{figure}

\begin{figure}
	\includegraphics[width=80mm]{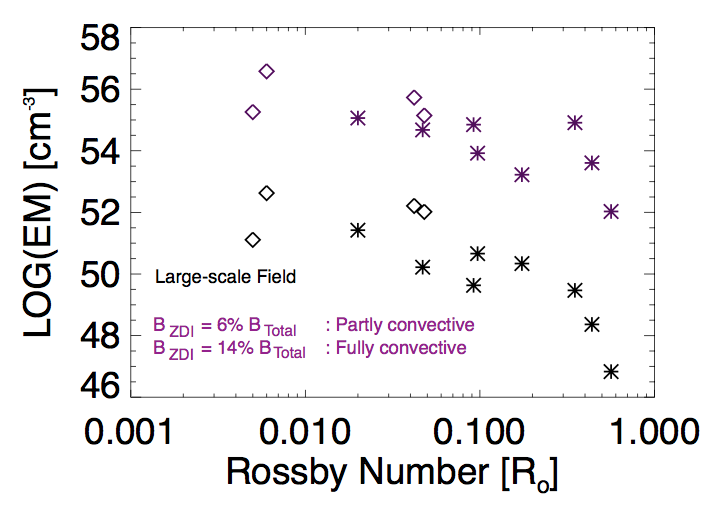}
\caption{X-ray Emission Measure as a function of Rossby number for both the large-scale field (black symbols) and the simulated small-scale + large-scale field (purple symbols).  \textit{Symbols}: asterisks represent partly convective dwarfs, $M > 0.4M_{\odot}$, and diamond represent fully convective dwarfs, $M \le 0.4M_{\odot}$.
	 \label{fig.EM_RossbyRatio6_Early_Late_2} }
\end{figure}

The X-ray luminosity has been shown to correlate well with either rotational velocity or Rossby number Ro (the ratio of the stellar rotation period, P, to the convective turnover time, $\tau_{c}$)  (e.g., \citealt{Pizzolato_ActivityRotation_2003,Pizzolato_StellarActivity_2003,Jeffries_coronalSaturation_2011}); in general, $L_{X}/L_{bol}$ increases and then saturates ($L_{X}/L_{bol} \approx10^{-3}$; \citep{Delfosse_LxLbolM7_1998}) with increasing rotation rate.  This behaviour is attributed to coronal saturation \citep{Vilhu_Chromospheric_1987,Stauffer_radial_1994} and occurs at Ro $\approx$ 0.1.  For a sample of M dwarfs, \citet{Lang_large-scale_2012}) reproduce the saturation of the X-ray emission measure for the large-scale field detected through ZDI.  These results are shown in Figs. (\ref{fig.EM_RossbyRatio6_Early_Late}) and (\ref{fig.EM_RossbyRatio6_Early_Late_2}) as black symbols.  

With the addition of small-scale field, the correlation between the X-ray emission measure and Rossby number changes depending on the amount of small-scale flux added.  For case (1), shown in Fig. (\ref{fig.EM_RossbyRatio6_Early_Late}), the rise and saturation of the X-ray emission measure is not as prominent with the addition of 500G of small-scale flux (red symbols) and is no longer evident for 1000G of small-scale flux (blue symbols).  This means that adding the same surface distribution of small-scale field to each star destroys the relationship between magnetic flux and Rossby number.  If we now consider case (2) where $B_{\mathrm{ls}} = 6\%B_{\mathrm{Total}}$, for the partly-convective M dwarfs and $B_{\mathrm{ls}} = 14\%B_{\mathrm{Total}}$, for the fully convective M dwarfs shown in Fig.(\ref{fig.EM_RossbyRatio6_Early_Late_2}) the rise and saturation of the X-ray emission with Rossby number is once again apparent.  The magnitude of the X-ray emission has increased by approximately 5 orders of magnitude due to the increase in flux (Table \ref{tab.stellardata}).  

For comparison with our previous work \citep{Lang_large-scale_2012} conducted on the field visible only to ZDI, we keep the model parameters e.g., the temperature ($T = 2\times10^{6}$K), source surface ($R_{ss} = 2.5R_{*}$) and $\kappa = 10^{-6}$, constant.  Keeping $\kappa$ constant results in an increase in pressure and coronal density due to the increase in the surface flux.  The coronal densities (shown in Table \ref{tab.stellardata}) now lie at the higher end of the accepted range: $10^{9}-10^{12}\mathrm{cm^{-3}}$ \citep{Ness_Density_ADLeo_2002,Ness_Density_2004}, as opposed to their previous values which were at the lower end.  The value of $\kappa$ could be altered in such a way to reduce the coronal densities back to the values calculated for the ZDI maps and in turn reduce the magnitude of the X-ray emission measure.

Comparison of the results of Fig. (\ref{fig.EM_RossbyRatio6_Early_Late}) and Fig. (\ref{fig.EM_RossbyRatio6_Early_Late_2}) indicate that the addition of the same small-scale field to each star removes the activity-rotation relation; however, scaling the small-scale field to the large-scale, e.g., $B_{\mathrm{ls}} = 6\%B_{\mathrm{Total}}$, for the partly-convective M dwarfs and $B_{\mathrm{ls}} = 14\%B_{\mathrm{Total}}$, for the fully convective M dwarfs, recovers the relation.  This would suggest that the small-scale field has the same dependence on rotation period as the large-scale field.  

In our previous work \citep{Lang_large-scale_2012}, the rotational modulation of the X-ray emission measure for the stellar sample did not demonstrate any trend with Rossby number and could not be used as an indicator of field topology due to too many contributing factors e.g., the angle of stellar inclination and the angle of the magnetic dipole axis from the rotation pole.  We find that with the addition of small-scale field, both with the same surface distribution and magnitude, and scaled with respect to the large-scale field, this is still the case.

With the addition of small-scale field, the magnitude of the rotation modulation of the X-ray Emission Measure changes for each star as a result of changes in surface field strength (See Table (\ref{tab.stellardata})).  For OT Ser, an early-M Dwarf, the change in the rotational modulation between the large-scale field and the addition of small-scale flux is nearly 30$\%$, whereas for GJ 49, also an early-M dwarf, the change in the modulation is 1$\%$.  Since we do not find any great change in the large-scale field structure with the addition of small-scale field, the change in the magnitude of the rotational modulation could be a result of the small-scale field producing low-lying small closed field regions which carpet the stellar surface including the areas where the large-scale field is open.

\subsection{Open Flux and Spin Down}\label{sec.FluxComparison}

\begin{figure*}
	\begin{center}
	\subfloat a){\label{fig.flux-a}}\includegraphics[width=80mm]{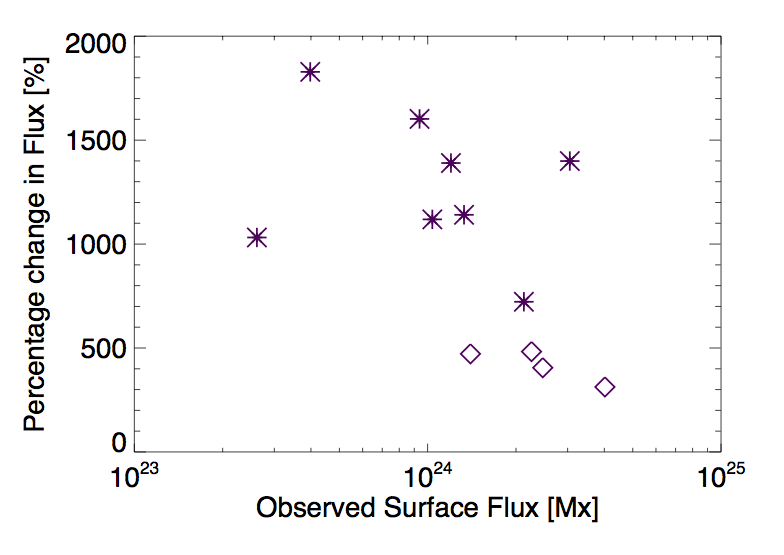}
	\subfloat b){\label{fig.open-b}}\includegraphics[width=80mm]{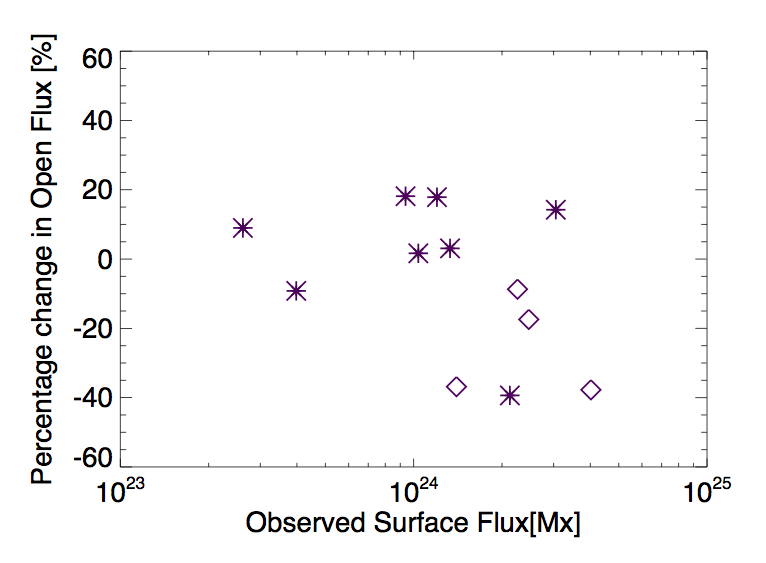}
\caption{The percentage change in (a) the surface flux and (b) the open flux as a function of the observed surface flux, due to the addition of small-scale field.  Symbols are as of Fig \ref{fig.EM_RossbyRatio6_Early_Late}.
	 \label{fig.Smallscale_openflux_comp} }
	 \end{center}
\end{figure*}

\begin{figure}
	\includegraphics[width=85mm]{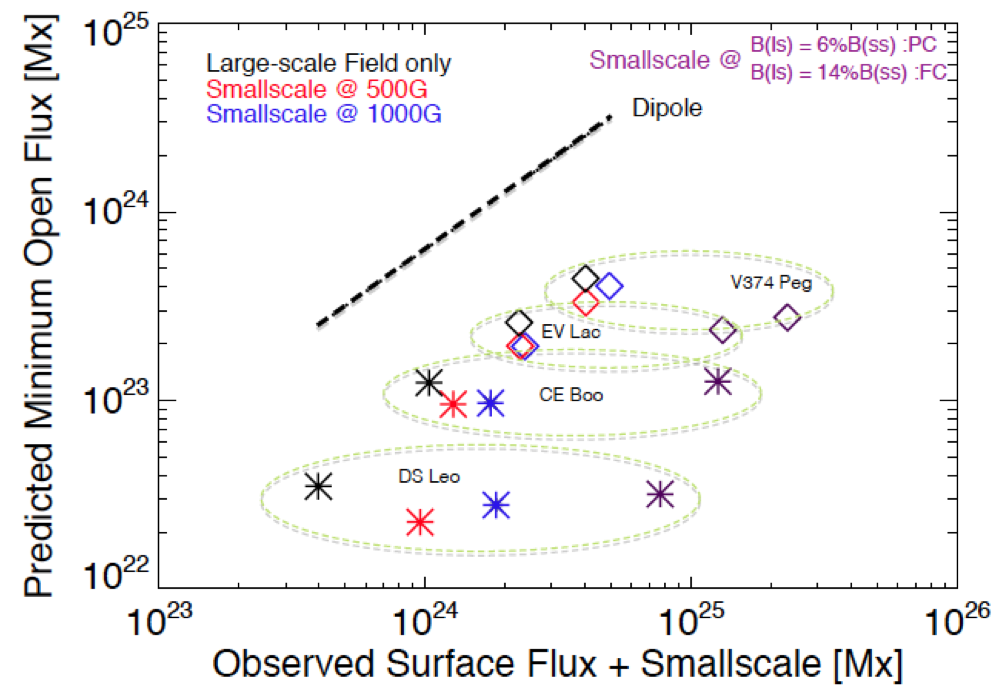}
\caption{Comparison of the magnitude of the minimum predicted open flux as a function of the observed surface flux with and without small-scale field.  The dashed line shows the predicted open flux of a pure dipole.  Four stars which span the spectral range of our sample have been chosen to show that when producing a model for the stellar corona a range of \textit{l} and \textit{m} modes must be considered to reproduce the correct coronal structure.  Symbols are as of Fig \ref{fig.EM_RossbyRatio6_Early_Late}.
	 \label{fig.Smallscale_fieldstrength_comp_2} }
\end{figure}

Coronal structure is important as it determines the X-ray emission from regions of closed magnetic field but also areas where the magnetic field is open and the stellar wind forms.  For stars with weaker large-scale magnetic fields, the range of field strengths present on the star has been altered by the addition of small-scale field.  This has little effect on the geometry of the large-scale field (Fig. \ref {fig.surface_get_1}).  The dipole axis and the location and extent of the open field regions are largely unchanged.  This is to be expected as the small-scale field is distributed axisymmetrically over the surface so it has no preferred direction.  This suggests that the latitudes from which a stellar wind could be launched would not be affected by the presence of small-scale field.  Within regions where the large-scale field is open there may still be a carpet of small-scale field, which could contribute to powering the stellar wind (e.g., \citet{Nishizuka_AnemoneJets_2011}).

The stellar wind is responsible for angular momentum loss and influences the stellar spin down time.  To investigate the effect the small-scale field has on the overall coronal structure, we examine both its influence on the total magnetic flux at the surface of the star and also the total open flux.  We analyse the geometry of the field by predicting and comparing the open flux to observed surface flux values obtained from the overall combination of \textit{l} and \textit{ m } modes, 
\begin{equation}
\frac{\Phi_{\mathrm{Open}}}{\Phi_{\mathrm{Surface}}}=\frac{R_{ss}^{2}\int|B_{r}(R_{ss},\theta,\phi)|\mathrm{d}\Omega}{R_{*}^{2}\int|B_{r}(R_{*},\theta,\phi)|\mathrm{d}\Omega}   \quad ,
\label{eq.field_8}
\end{equation}
where $\Omega$ is the solid angle.  We note this gives a lower limit to the true open flux as on some field lines the gas pressure may exceed the magnetic pressure.

Adding in small-scale field increases the surface flux.  Fig.(\ref{fig.Smallscale_openflux_comp}a) shows this increase expressed as 
\begin{equation}
\Delta \phi_{\mathrm{Surface}} = \frac{(\phi_{\mathrm{Surface}})_{\mathrm{large + small}}}{(\phi_{\mathrm{Surface}})_{\mathrm{large}}} - 1 \quad .
\end{equation}
The fractional increase in surface flux is clearly greatest for those stars whose large-scale surface flux is lowest.

The addition of the small-scale field also results in a slight change (10$\%$ to 40$\%$) in the open flux (Fig.\ref{fig.Smallscale_openflux_comp}b) but shows no preference between partly-convective and fully-convective stars..  The fraction for the open flux is given by
\begin{equation}
\Delta \phi_{\mathrm{Open}} = \frac{(\phi_{\mathrm{Open}})_{\mathrm{large + small}}}{(\phi_{\mathrm{Open}})_{\mathrm{large}}} - 1 \quad.
\end{equation}

This result is in keeping with the increase in surface flux, which would increase the magnetic pressure (Eq. \ref{eq.pressure_1}).  We note that the magnitude of the open flux is dependent on the chosen value for the source surface (as $R_{ss} =\rightarrow \infty$, $\Phi_{\mathrm{Open}}\rightarrow0$) but the effect of adding in small-scale field is the same for all values of the source surface.  As there is little change in the open flux with the addition of small-scale field, we would not expect the angular momentum or mass loss to be significantly affected.

For many stars a full surface magnetic map is not available and only a single flux estimate is possible. Assuming that all of the surface flux is contained in one single mode, for example a dipole, can however lead to an overestimate of the amount of open flux.  As discussed in \citet{Lang_large-scale_2012}, the open flux for any single mode is simply related to the surface flux as:
\begin{equation}
\frac{\Phi_{\mathrm{Open}}}{\Phi_{\mathrm{Surface}}} = \frac{(2l + 1)(\frac{R_{ss}}{R_*})^{l+1} }{l + (l + 1)(\frac{R_{ss}}{R_*})^{2l + 1}  }       \quad .
\end{equation}
Figure (\ref{fig.Smallscale_fieldstrength_comp_2}) shows that when small-scale field is added there is an increase in surface flux but the predicted open flux is still at least an order of magnitude smaller than it would be had we only considered the dipole modes.  Therefore, the angular momentum loss, $\dot{J}$, due to the stellar wind, which is determined by the amount of open flux i.e. a Weber-Davis model \citep{Weber_Davis_Model_1967}, given by
\begin{equation}
\dot{J} \propto \Phi_{\mathrm{Open}}^{2}    ,
\label{eq.WeberDavis}
\end{equation}
is influenced by the topology of the field and would be overestimated by at least 2 orders of magnitude.  This is also true for the mass loss rate, 
\begin{equation}
\dot{M} \propto \Phi_{\mathrm{Open}}    ,
\label{eq.MassLoss}
\end{equation}
which could be overestimated by an order of magnitude if the topology is over-simplified.  We conclude from this result that when producing a model for the stellar corona, or the stellar wind, a range of \textit{l} and \textit{m} modes must be considered to reproduce the correct, more complex, coronal structure.

\begin{table*}
\begin{center}
\caption{Values for mass, radius, Rossby number and $\left< \mathrm{B_{V}}\right >$ (the average large-scale magnetic flux derived from spectropolarimetric measurements), are from \citet{Donati_EarlyM_2008,Morin_MidM_2008}.  Values of $\frac{\left< \mathrm{B_{V}}\right >}{\left< \mathrm{B_{I}}\right >}$ for  GJ 182, DT Vir, Ce Boo, AD Leo, EV Lac and YZ CMi are from \citet{Reiners_Basri_MagneticTopology_2009}, while values for DS Leo, GJ 49, OT Ser EQ Peg A, V374 Peg, and EQ Peg B are estimates (depicted by $^{e}$) based on \citet{Reiners_Basri_MagneticTopology_2009}.  Values for $\left<\mathrm{B_{r}^{ls}}\right >$ (the average large-scale [ls] radial flux), $\left<\mathrm{B_{ss}}\right >$ (the average small-scale [ss] flux, scaled with respect to the large-scale field) and $\left<\mathrm{B_{ls + ss}}\right >$ (the average large- + small-scale flux) as well as $\phi_{Open}$ (the open flux) and $\phi_{Surface}$ (the surface flux) values are from this work.
\label{tab.stellardata_dipole_contribution}
}
\centering
\begin{tabular}{!{\color{black}\vrule}r!{\color{black}\vrule}cc!{\color{black}\vrule}cc!{\color{black}\vrule}cc!{\color{black}\vrule}c!{\color{black}\vrule}cc!{\color{black}\vrule}cc!{\color{black}\vrule}cc!{\color{black}\vrule}cc!{\color{black}\vrule}}
\hline
\multirow{3}{*} & & & & & 500G & 1000G & & & & & & & & & \\
\hline
{Star} & Mass & Ro & $\left<B_{\mathrm{V}}\right >$ & $\left<B_{r}^{\mathrm{ls}}\right >$ & $\left<B_{r}^{\mathrm{ls + ss}}\right >$ & $\left<B_{r}^{\mathrm{ls + ss}}\right >$ & $\frac{\left<B_{\mathrm{V}}\right >}{\left< B_{\mathrm{I}}\right >}$ & $\left< B_{\mathrm{ss}}\right >$ & $\left< B_{R}^{\mathrm{ls + ss}}\right >$ & $|\beta_{\mathrm{M}}^{\mathrm{ls}}|$ &  $|\beta_{\mathrm{M}}^{\mathrm{ls}}|$ & $\phi_{\mathrm{Open}}^{\mathrm{ls}}$ & $\phi_{\mathrm{Open}}^{\mathrm{ls + ss}}$ & $\phi_{\mathrm{Surface}}^{\mathrm{ls}}$ & $\phi_{\mathrm{Surface}}^{\mathrm{ls + ss}}$\\
&($\mathrm{M_\odot}$) & ($10^{-2}$) & (kG)  & (kG) & (kG) & (kG) & ($\%$) & (kG) & (kG) & ($^{\circ}$) & ($^{\circ}$) & 10$^{23}$ Mx & 10$^{23}$ Mx & 10$^{25}$Mx & 10$^{25}$Mx\\
\hline
\hline
GJ 182\quad(07) & 0.75  & 17.4 & 0.17 & 0.10 & 0.19 &  0.21 & 6 & 1.8 & 1.8 &  41.1 & 41.4  & 2.9 & 3.3 & 3.0 & 4.6\\
\hline
\multirow{2}{*}{DT Vir}\quad(07) & 0.59 & 9.2 & 0.15 & 0.08 & 0.18 & 0.20 & 5 &1.5 & 1.5 & 83.6 & 84.0 & 1.0 & 1.2 & 0.1 & 7.4   \\
\quad \qquad \quad(08) & - & - & 0.15 & 0.12 & 0.21 & 0.22 & 5 & 2.3 & 2.3 &20.0 & 13.0 & 0.5 & 1.2 & 0.1 &2.3 \\
\hline
\multirow{2}{*}{DS Leo}\quad(07) & 0.58 & 43.8 & 0.10 & 0.04 & 0.16 & 0.18 & 5$^{e}$ &0.75 & 0.76 & 41.1 & 38.8 & 0.5 & 0.3 & 0.05 & 0.8  \\
\quad \qquad  \quad(08) & - & - & 0.9 & 0.03 & 0.16 & 0.18 & 5$^{e}$ & 0.60 & 0.3 & 42.9 & 42.6 & 0.3 & 0.3 & 0.04 & 0.6  \\
\hline
GJ 49\quad(07) & 0.57 & 56.4 & 0.3 & 0.02 &0.15 & 0.18 &6$^{e}$ & 0.30 & 0.30 & 10.9 & 2.3 & 0.30 & 0.30 & 0.03 & 0.30  \\
\hline
OT Ser\quad(08) & 0.55 & 9.70 & 0.14 & 0.13 & 0.19 & 0.22 & 6$^{e}$ & 1.9 & 0.8 &12.1 & 12.8 & 1.0 & 4.8 & 0.09 & 4.0   \\
\hline
CE Boo\quad(08) & 0.48 & 35.0 & 0.10 & 0.12 & 0.20 & 0.23 & 6 & 1.8 & 1.8 & 7.4 & 6.3 & 1.2 & 1.3 & 0.1 & 1.3  \\
\hline
\multirow{2}{*}{AD Leo}\quad(07) &0.42 & 4.7 & 0.19 & 0.21 & 0.27 & 0.30 & 7& 2.8 & 2.9 & 4.5 & 4.5 & 1.5 & 1.6 & 0.1 &1.6  \\
\quad \qquad \quad(08) & - & - & 0.18 & 0.21 & 0.27 & 0.29 & 7 & 2.8 & 2.9 & 7.3 & 7.3 & 1.5 & 1.6 & 0.1 & 1.6 \\
\hline
EQ Peg A\quad(06) & 0.39 & 2.0 & 0.48 & 0.39 & 0.42 & 0.44 & 10$^{e}$& 3.8 & 3.8 & 25.9 & 25.9 & 2.4 & 1.5 & 0.2 & 1.7  \\
\hline
\multirow{2}{*}{EV Lac}\quad(06) & 0.32 & 6.8 & 0.57 & 0.63 & 0.66 & 0.67 & 13 & 3.8 & 3.9 & 45.8 & 45.8 & 2.8 & 2.4 & 0.2 & 1.3  \\
\quad \qquad \quad(07) & - & - & 0.49 & 0.57 & 0.59 & 0.61 & 13 & 3.8 & 3.9 &43.2 &43.3 & 2.6 & 1.7 & 0.2 & 0.96  \\
\hline
\multirow{2}{*}{YZ CMi}\quad(07) &0.31 & 4.2 & 0.56  & 0.73 & 0.74 & 0.76 & 14& 3.8 & 3.9 &24.8 & 24.8 & 3.1 & 2.4 & 0.3 & 1.3  \\
\quad \qquad \quad(08) & - & - & 0.55 & 0.66 & 0.69 & 0.70 & 14 & 3.9 & 4.0 & 12.1 & 11.1 & 3.0 & 1.8 & 0.3 & 0.92  \\
\hline
V374 Peg\quad(05) & 0.28 & 0.6 & 0.78 & 1.0 & 1.1 & 1.1 & 14$^{e}$& 5.9 & 6.0 & 9.3 & 8.0 & 4.4 & 2.0 & 0.4 & 1.7 \\
\hline
EQ Peg B\quad(06) & 0.25 & 0.5 & 0.45 & 0.49 & 0.51 & 0. 53 & 14$^{e}$ & 3.3 & 3.4 & 6.5 & 6.4 & 1.7 & 1.0 & 0.1 & 0.8 \\
\hline
\end{tabular}
\end{center}
\end{table*}

\section{Summary}\label{sec.Summary}

We have created a model for small-scale field using synthesised spot distribution maps.  We allocate field strengths in a Gaussian distribution from the centre of the spot by either (1) fixing the value of $B_{\mathrm{max}}$ to be either $\pm$500G or $\pm$1000G; or (2) setting the value of $B_{\mathrm{max}}$ such that the large-scale field contributes only 6$\%$ of the total field for partly-convective M dwarfs and 14$\%$ of the total field for fully-convective M dwarfs, as indicated in \citet{Reiners_Basri_MagneticTopology_2009}.  We have incorporated the radial surface map produced by this model into the reconstructed maps of the observed radial magnetic field at the stellar surface for a sample of early-to-mid M dwarfs and extrapolated their 3D coronal magnetic field using the Potential Field Source Surface method.  

We have investigated the effect the addition of small-scale field has on the topology of the large-scale magnetic field at the stellar surface and the structure of the extrapolated 3D corona.  By assuming a hydrostatic, isothermal corona, we have determined the following:\\
1.  The geometry of the magnetic field e.g., the angle of the dipole axis, overall large-scale structure of the 3D extrapolated corona and location of coronal holes where the stellar wind is emitted all remain largely unchanged.\\
2.  Addition of the same small-scale field to each star removes the $L_{X}-Ro$ relation; however, scaling the small-scale field to the large-scale (ZDI) field recovers the relation.  We conclude from this that the small-scale field has the same dependence on rotation period as the large-scale field.\\
3.  The magnitude of the rotational modulation of the X-ray emission measure changes with the addition of more surface flux; however, no trend with Rossby number Ro is found.  This change could be due to the carpet of low-lying field.\\
4.  The addition of small-scale field increases the surface flux.\\
5.  And finally, we find that the large-scale open flux does not vary greatly with the addition of small-scale field.  This suggests that the mass loss rate, the angular momentum loss and the spin down time for a star, are not significantly affected by small-scale flux.\\


\section*{Acknowledgements}
PL acknowledges support from an STFC studentship.  JM, AV and RF acknowledge support from fellowships of the Alexander von Humboldt foundation, the Royal Astronomical Society and STFC, respectively.  The authors would like to thank the referee for a particularly thorough and detailed report.

\bibliography{MDwarfs.bib}
\bsp
\label{lastpage}
\end{document}